\documentclass[12pt]{article}
\usepackage{amsfonts, amsmath, amssymb, cite, graphicx, }

\usepackage{color}
\usepackage[all, knot]{xy}
\usepackage{tikz}
\usepackage{braket}

\usepackage[utf8]{inputenc}
\usepackage{ulem}

\usepackage{epstopdf}
\usepackage[footnotesize]{caption}
\usepackage{amsthm}
\usepackage{enumitem}
\usepackage{mathrsfs}

\usepackage[margin=3cm]{geometry}

\def \be {\begin{equation}}
\def \ee {\end{equation}}
\def \ba {\begin{aligned}}
\def \ea {\end{aligned}}
\def \bea {\begin{eqnarray}}
\def \eea {\end{eqnarray}}

\begin{document}

\begin{titlepage}
\vspace{0.5cm}
\begin{center}
{\Large \bf Correlation functions, entanglement and chaos in the $T\bar{T}$/$J\bar{T}${-}deformed CFTs}

\lineskip .75em
\vskip 2.5cm
{\large Song He$^{a,b,}$\footnote{hesong@jlu.edu.cn}, Hongfei Shu$^{c,}$\footnote{shuphy124@gmail.com } }
\vskip 2.5em
 {\normalsize\it $^{a}$Center for Theoretical Physics and College of Physics, Jilin University, Changchun 130012, People's Republic of China\\
 $^{b}$Max Planck Institute for Gravitational Physics (Albert Einstein Institute),\\
Am M\"uhlenberg 1, 14476 Golm, Germany\\
$^{c}$ Department of Physics, Tokyo Institute of Technology
Tokyo, 152-8551, Japan}
\vskip 3.0em
\end{center}
\begin{abstract}
In this paper, we regard the $T\bar{T}$/$J\bar{T}${-}deformed CFTs as perturbation theories and calculate the first order correction of the correlation functions due to the $T\bar{T}$/$J\bar{T}${-}deformation. As applications, we study the R\'enyi entanglement entropy of excited state in the $T\bar{T}$/$J\bar{T}${-}deformed two-dimensional CFTs. We find, up to the first order perturbation of the deformation, the R\'enyi entanglement entropy of locally excited states will acquire a non-trivial time dependence. The excess of the R\'enyi entanglement entropy of locally excited state is changed up to order ${\cal O}(c)$. Furthermore, the out of time ordered correlation function is investigated to confirm that the $T\bar{T}$/$J\bar{T}$-deformations do not change the maximal chaotic behavior of holographic CFTs up to the first order of the deformations.
  \end{abstract}
\end{titlepage}

\baselineskip=0.7cm

\tableofcontents
\newpage
\section{Introduction}
Recently much attention has been paid to the irrelevant deformations of two-dimensional quantum field theories by bilinear form of conserved currents \cite{Smirnov:2016lqw}. This deformation
\cite{Zamolodchikov:2004ce, Cavaglia:2016oda} has been extensively investigated \cite{Dubovsky:2012wk,Caselle:2013dra,Dubovsky:2017cnj,
Cardy:2018sdv,Aharony:2018vux, Chakraborty:2018kpr,Datta:2018thy,Aharony:2018bad,Chen:2018eqk,Conti:2018tca,Bonelli:2018kik,Lashkevich:2018jmo,Chang:2018dge,Baggio:2018rpv,
Dubovsky:2018bmo, Conti:2018jho,  Chen:2018keo, Santilli:2018xux, Jiang:2019tcq, LeFloch:2019rut, Jiang:2019hux, Conti:2019dxg, Chang:2019kiu}. Although these deformations are  irrelevant in the renormalization group sense, the deformed theory appears to be more predictive than the generic non-renormalizable QFT. Remarkably, such deformation preserves the integrability for the integrable quantum field \cite{Smirnov:2016lqw}. Even for the non-integrable theory, some properties, e.g. for a finite size spectrum and the S-matrix, of the $T\bar{T}/J\bar{T}$-deformed theory can be exactly calculated based on the data of the undeformed theory \cite{Smirnov:2016lqw, Cavaglia:2016oda}. For 
 Non-Lorentz invariant cases were studied in \cite{Guica:2017lia, Chakraborty:2018vja, Aharony:2018ics, Cardy:2018jho, Nakayama:2018ujt,  Guica:2019vnb}.

$T\bar{T}$/$J\bar{T}${-}deformed CFTs have also been applied to string theory \cite{Giveon:2017nie,Giveon:2017myj,Asrat:2017tzd,Giribet:2017imm, Baggio:2018gct, Apolo:2018qpq, Babaro:2018cmq, Chakraborty:2018aji, Araujo:2018rho, Giveon:2019fgr, Chakraborty:2019mdf, Nakayama:2019mvq}, where the application to AdS/CFT is especially interesting. The holographic dual of the positive sign $T\bar{T}${-}deformed two-dimensional holographic CFT was proposed to be AdS$_3$ gravity with a finite radius cut-off in \cite{McGough:2016lol}. To check this proposal, various aspects, such as the energy spectrum and the propagation speed, must agree on both sides \cite{McGough:2016lol}. Some recent progress in regard to the holographic aspect of the deformations has been reported in \cite{Shyam:2017znq, Kraus:2018xrn, Cottrell:2018skz, Bzowski:2018pcy, Taylor:2018xcy, Hartman:2018tkw, Shyam:2018sro, Caputa:2019pam, Gorbenko:2018oov,Apolo:2019yfj}.

The R\'enyi  entanglement entropy has been used as a helpful quantity to measure the properties of the vacuum and excited states \cite{Calabrese:2004eu, Kitaev:2005dm, Levin:2006zz, Nozaki:2014hna, He:2014mwa,Guo:2015uwa, Chen:2015usa, He:2017lrg,He:2017vyf,He:2017txy,Guo:2018lqq, Apolo:2018oqv}. In the un-deformed CFTs, the $n$th R\'enyi entanglement entropy $S^{(n)}_A$ has been extensively studied in the literature \cite{Nozaki:2014hna,He:2014mwa,Caputa:2014vaa,Caputa:2014eta,Guo:2015uwa,Chen:2015usa, He:2017lrg,Jahn:2017xsg, Guo:2018lqq, Miyaji:2018atq,Kusuki:2018wpa,Shimaji:2018czt, Apolo:2018oqv,Kusuki:2019gjs,Caputa:2019avh}. In rational CFTs, it has been shown that the excess of the R\'enyi entanglement entropy has to be the logarithm of the quantum dimension \cite{He:2014mwa} of the corresponding local operator. The quantum entanglement of $T\bar{T}${-}deformed CFTs have been investigated in \cite{Chakraborty:2018kpr, Donnelly:2018bef,Chen:2018eqk,Gorbenko:2018oov,Park:2018snf,Banerjee:2019ewu,Murdia:2019fax,Ota:2019yfe, Sun:2019ijq, Jeong:2019ylz}. However, these works are mainly focused on the R\'enyi entanglement entropy of the vacuum states in deformed CFTs.  In this paper focus on the locally excited states in deformed CFTs. To initiate the study, we have to know the correlation function of CFTs\footnote{Recently,  \cite{Cardy:2019qao} have used the local and non-local field renormalizations to study the correlation functions which are shown to be UV finite to all orders formally.}.

Here, we study the correlation functions of primary operators in the $T\bar{T}$/$J\bar{T}${-}deformed two-dimensional CFTs without the effect of the renormalization group flow of the operator. We will focus on the 2- and 4-point functions in $T\bar{T}$/$J\bar{T}$. For simplicity, we regard the $T\bar{T}$/$J\bar{T}${-}deformed CFTs as perturbation theories of CFTs and compute the first order correction of the correlation function due to the $T\bar{T}$/$J\bar{T}${-}deformation. Since $T\bar{T}$/$J\bar{T}$ are conserved quantities, the Ward identities are held in the deformed theory. Once we implement the Ward identity for deformed correlation functions, we have to deal with the divergences. We apply the dimensional regularization procedure and the deformation of correlation function up to the first order can be obtained explicitly.

As applications, we employ the formula to investigate the R\'enyi entanglement entropy in two-dimensional CFTs perturbatively. We just put a local primary operator following the procedure in \cite{He:2014mwa, Guo:2015uwa, Chen:2015usa, He:2017lrg, Guo:2018lqq, Apolo:2018oqv} to obtain a locally excited state. With time evolution, the excess of R\'enyi entanglement entropy of deformed two-dimensional CFTs has been calculated in this paper. We will show how the $T\bar{T}$/$J\bar{T}$-deformation changes the excess of R\'enyi entanglement entropy in this local quenched system.

The $T\bar{T}${-}deformation of the integrable model is proposed to hold the integrability structure. Alternatively, we would like to employ the out of time order correlation function (OTOC) \cite{Shenker:2014cwa, Maldacena:2015waa, Roberts:2014ifa} to gain some insight into integrability/chaos after the deformation, since the OTOC has been broadly regarded as one of the quantities to capture the chaotic or integrable. We investigate the OTOC in the deformed CFTs to see whether the chaotic property is preserved or not after the $T\bar{T}$/$J\bar{T}${-}deformation in a perturbative sense.

The remainder of this paper is organized as follows.
In Section 2, we setup the perturbation of $T\bar{T}${-}deformed theories.
In terms of perturbative CFT technicals, we formulate 2- and 4-point correlation functions of $T\bar{T}$-deformed CFTs explicitly, where the dimensional regularization has been implemented. In Section 3, we have studied the excess of  the R\'enyi entanglement entropy of the locally excited states in $T\bar{T}${-}deformed theory. In Section 4, we work out the out of time ordered correlation function of the $T\bar{T}${-}deformed theory, up to first order perturbation. In Section 5, we directly extend the investigations in Sections 2, 3 and 4 to the $J\bar{T}${-}deformed theory. Finally, section 6 is devoted to conclusions and discussions. We also mention some likely future problems. In the appendices, we would like to list some techniques and notations relevant to our analysis.


\section{$T\bar{T}$-deformation and correlation functions}
In this section, we give a lightning review of the $T\bar{T}$-deformation and calculate the correlation function in the $T\bar{T}$-deformed CFTs, which are useful in the later parts.

The $T\bar{T}${-}deformed action is the trajectory on the space of field theory satisfying
\be
\frac{dS(\lambda)}{d\lambda}=\int d^2z\sqrt{g}(T\bar{T})_\lambda,
\ee
where $\lambda$ is the coupling constant of the $T\bar{T}$-operator. $S(\lambda=0)$ is the action of the un-deformed CFT on the flat metric $ds^2=dzd\bar{z}$. Since the theory is on the flat space, the $T\bar{T}$ operator can be written as
\be
T\bar{T}=T_{zz}T_{\bar{z}\bar{z}}-T_{z\bar{z}}T_{z\bar{z}}.
\ee
with $T=T_{zz}$ and $\bar{T}=T_{\bar{z}\bar{z}}$. In this paper, we will focus on the perturbation theory of $S(\lambda)$, i.e.
\be
S(\lambda)=S(\lambda=0)+\lambda\int d^2z\sqrt{g}(T\bar{T})_{\lambda=0}+{\cal O}(\lambda^2),
\ee
where $(T\bar{T})_{\lambda=0}=T\bar{T}$ plays the role of the perturbation operator in the CFT. Without confusion we will denote $(T\bar{T})_{\lambda=0}$ as $T\bar{T}$ from now on.

In this perturbation theory, the first order correction to the $n$-point correlation function of primary operators $\langle {\cal O}_1(z_1,\bar{z}_1){\cal O}_2(z_2,\bar{z}_2)\cdots {\cal O}_n(z_n,\bar{z}_n) \rangle$
 becomes
\be
\langle {\cal O}_1(z_1,\bar{z}_1){\cal O}_2(z_2,\bar{z}_2)\cdots {\cal O}_n(z_n,\bar{z}_n) \rangle_{\lambda}=\lambda\int d^2z\langle T\bar{T}(z,\bar{z}){\cal O}_1(z_1,\bar{z}_1){\cal O}_2(z_2,\bar{z}_2)\cdots {\cal O}_n(z_n,\bar{z}_n) \rangle.
\ee
By using the Ward identity, this correction can be written as
\be
\ba
\langle {\cal O}_1(z_1,\bar{z}_1){\cal O}_2(z_2,\bar{z}_2)\cdots {\cal O}_n(z_n,\bar{z}_n) \rangle_{\lambda}
=&
\lambda\int d^2z \Big(\sum_{i=1}^{n}\big(\frac{h_{i}}{(z-z_{i})^{2}}+\frac{\partial_{z_{i}}}{z-z_{i}}\big)\Big)\Big(\sum_{i=1}^{n}\big(\frac{\bar{h}_{i}}{(\bar{z}-\bar{z}_{i})^{2}}+\frac{\partial_{\bar{z}_{i}}}{\bar{z}-\bar{z}_{i}}\big)\Big)\\&\qquad \times \langle {\cal O}_1(z_1,\bar{z}_1){\cal O}_2(z_2,\bar{z}_2)\cdots {\cal O}_n(z_n,\bar{z}_n) \rangle,
\ea
\ee
where we have used the fact that any correlation function including $T_{z\bar{z}}$ vanish i.e. $\langle T_{z\bar{z}} \cdots \rangle=0$. As examples, we will study the corrections to the 2- and 4-point functions up to the first order perturbation in $\lambda$.

Let us first consider the two-point function of primary operator:
\be
\langle{\cal O}(z_{1}, \bar{z}_{1}){\cal O}(z_{2}, \bar{z}_{2})\rangle=\frac{C_{12}}{z_{12}^{2h}\bar{z}_{12}^{2\bar{h}}},
\ee
where $z_{ij}=z_i-z_j, \bar{z}_{ij}=\bar{z}_i-\bar{z}_j$. The Ward identity leads to
\be
\ba
\label{eq:TTbarOO}
&\quad \langle T(z)\bar{T}(z){\cal O}(z_{1}, \bar{z}_{1}){\cal O}(z_{2}, \bar{z}_{2})\rangle\\
&=\Big(\frac{h\bar{h}z_{12}^{2}\bar{z}_{12}^{2}}{(z-z_{1})^{2}(z-z_{2})^{2}(\bar{z}-\bar{z}_{1})^{2}(\bar{z}-\bar{z}_{2})^{2}}{-\frac{4\pi\bar{h}\bar{z}_{12}^{2}}{(z-z_{1})(\bar{z}-\bar{z}_{1})(\bar{z}-\bar{z}_{2})^{2}}\delta^{(2)}(z-z_{1}) }\\
&\quad {-\frac{4\pi\bar{h}\bar{z}_{12}^{2}}{(z-z_{2})(\bar{z}-\bar{z}_{1})^{2}(\bar{z}-\bar{z}_{2})}\delta^{(2)}(z-z_{2})\Big) } \langle{\cal O}(z_{1}){\cal O}(z_{2})\rangle.
\ea
\ee
Note that due to the effect of  $\partial_{z_i}$,  the terms such as $(\bar{z}-\bar{z}_i)^{-1}$ will provide a delta function. One can see that the final two terms will contribute to two non-dynamics terms which are UV divergence after the space time integration. For simplicity, we drop out these terms in later analysis. One can show that the first term in above equation is consistent with the two-point function given by \cite{Kraus:2018xrn, Cardy:2019qao}. Using the formula (\ref{eq:I2-regulated}) in appendix, we obtain the first order correction of the two-point function due to the $T\bar{T}${-}deformation
\be
\ba
\label{2pt-regular}
\langle T(z)\bar{T}(z){\cal O}(z_{1}, \bar{z}_1){\cal O}(z_{2}, \bar{z}_2)\rangle_{\lambda}=&\lambda h\bar{h}z_{12}^{2}\bar{z}_{12}^{2}{\cal I}_{2}(z_{1},z_{2}) \langle{\cal O}(z_{1}, \bar{z}_1){\cal O}(z_{2}, \bar{z}_2)\rangle
\\=&
\lambda h\bar{h}\frac{4\pi}{|z_{12}|^{2}}\Big(\frac{4}{\epsilon}+2\log(\mu|z_{12}|^{2})+2\log\pi+2\gamma-5\Big)\langle{\cal O}(z_{1}, \bar{z}_1){\cal O}(z_{2}, \bar{z}_2)\rangle,
\ea
\ee
where $\epsilon$ is the dimensional regularization parameter. See (\ref{eq:integrals}) for the notation of ${\cal I}_2$. This result reproduces the one obtained in \cite{Kraus:2018xrn, Cardy:2019qao}. In the following, we will use the same prescription to handle the divergent integrals.

For the single primary operator ${\cal O}$, we have
\be
\langle{\cal O}^\dagger(z_{1},\bar{z}_1){\cal O}(z_{2}, \bar{z}_2){\cal O}^\dagger(z_{3}, \bar{z}_3){\cal O}(z_{4}, \bar{z}_4)\rangle=\frac{G(\eta,\bar{\eta})}{z_{13}^{2h}z_{24}^{2h} \bar{z}_{13}^{2\bar{h}}\bar{z}_{24}^{2\bar{h}}},
\ee
with the cross ratios
\be
\eta=\frac{z_{12}z_{34}}{z_{13}z_{24}}, \quad \bar{\eta}=\frac{\bar{z}_{12}\bar{z}_{34}}{\bar{z}_{13}\bar{z}_{24}}.
\ee
The first order correction for this four-point function due to $T\bar{T}${-}deformation is
\be
\ba
&\langle{\cal O}^\dagger (z_{1},\bar{z}_{1}){\cal O}(z_{2},\bar{z}_{2}){\cal O}^{\dagger}(z_{3},\bar{z}_{3}){\cal O}(z_{4},\bar{z}_{4})\rangle_{\lambda}
\\=&
\lambda \int d^{2}z\Big\{\Big(\frac{hz_{13}^{2}}{(z-z_{1})^{2}(z-z_{3})^{2}}+\frac{hz_{24}^{2}}{(z-z_{2})^{2}(z-z_{4})^{2}}+\frac{z_{23}z_{14}}{\prod_{j=1}^{4}(z-z_{j})}\frac{\eta\partial_{\eta}G(\eta,\bar{\eta})}{G(\eta,\bar{\eta})}\Big)
\\&
\Big(\frac{\bar{h}\bar{z}_{13}^{2}}{(\bar{z}-\bar{z}_{1})^{2}(\bar{z}-\bar{z}_{3})^{2}}+\frac{\bar{h}\bar{z}_{24}^{2}}{(\bar{z}-\bar{z}_{2})^{2}(\bar{z}-\bar{z}_{4})^{2}}+\frac{\bar{z}_{23}\bar{z}_{14}}{\prod_{j=1}^{4}(\bar{z}-\bar{z}_{j})}\frac{\bar{\eta}\partial_{\bar{\eta}}G(\eta,\bar{\eta})}{G(\eta,\bar{\eta})}\Big)
\\&
-\eta\bar{\eta}\frac{z_{23}z_{14}\bar{z}_{23}\bar{z}_{14}}{\prod_{i}^{4}(z-z_{i})(\bar{z}-\bar{z}_{i})}\frac{\partial_{\eta}G(\eta,\bar{\eta})\partial_{\bar{\eta}}G(\eta,\bar{\eta})}{G(\eta,\bar{\eta})^{2}}+\eta\bar{\eta}\frac{z_{23}z_{14}\bar{z}_{23}\bar{z}_{14}}{\prod_{i}^{4}(z-z_{i})(\bar{z}-\bar{z}_{i})}\frac{\partial_{\eta}\partial_{\bar{\eta}}G(\eta,\bar{\eta})}{G(\eta,\bar{\eta})}\\
&-\frac{4\pi\bar{h}\bar{z}_{13}^{2}}{(z-z_{1})(\bar{z}-\bar{z}_{1})(\bar{z}-\bar{z}_{3})^{2}}\delta^{(2)}(z-z_{1})-\frac{4\pi\bar{h}\bar{z}_{13}^{2}}{(z-z_{3})(\bar{z}-\bar{z}_{3})(\bar{z}-\bar{z}_{1})^{2}}\delta^{(2)}(z-z_{3})\\&-\frac{4\pi\bar{h}\bar{z}_{24}^{2}}{(z-z_{2})(\bar{z}-\bar{z}_{2})(\bar{z}-\bar{z}_{4})^{2}}\delta^{(2)}(z-z_{2})-\frac{4\pi\bar{h}\bar{z}_{24}^{2}}{(z-z_{4})(\bar{z}-\bar{z}_{4})(\bar{z}-\bar{z}_{2})^{2}}\delta^{(2)}(z-z_{4})\\&-\frac{\bar{z}_{12}\bar{z}_{34}\partial_{\bar{\eta}}G(\eta,\bar{\eta})}{G(\eta,\bar{\eta})}\Big[\frac{2\pi\delta^{(2)}(z-z_{1})}{(z-z_{1})(\bar{z}-\bar{z}_{2})(\bar{z}-\bar{z}_{3})(\bar{z}-\bar{z}_{4})}+\frac{2\pi\delta^{(2)}(z-z_{2})}{(z-z_{2})(\bar{z}-\bar{z}_{1})(\bar{z}-\bar{z}_{3})(\bar{z}-\bar{z}_{4})}\\&+\frac{2\pi\delta^{(2)}(z-z_{3})}{(z-z_{3})(\bar{z}-\bar{z}_{1})(\bar{z}-\bar{z}_{2})(\bar{z}-\bar{z}_{4})}+\frac{2\pi\delta^{(2)}(z-z_{4})}{(z-z_{4})(\bar{z}-\bar{z}_{1})(\bar{z}-\bar{z}_{2})(\bar{z}-\bar{z}_{3})}\Big]
\Big\}\\
&\langle{\cal O}^{\dagger}(z_{1},\bar{z}_{1}){\cal O}(z_{2},\bar{z}_{2}){\cal O}^{\dagger}(z_{3},\bar{z}_{3}){\cal O}(z_{4},\bar{z}_{4})\rangle.
\ea
\ee
The delta functions presented in the 4-th and 5-th rows of the above equation will not contribute to the dynamics of the four-point function after the proper regularization, which is similar as the situation in two-point correlation function. We thus drop these terms. Moreover, after the space time integration associated with the deformation, the terms with  delta function and $\frac{\eta \partial_\eta G}{G}$ will also vanish.

Using the notations of the integrals introduced in (\ref{eq:integrals}), we express the first order correction to four-point function as
\be
\ba
&\langle{\cal O}(z_{1},\bar{z}_{1}){\cal O}(z_{2},\bar{z}_{2}){\cal O}^{\dagger}(z_{3},\bar{z}_{3}){\cal O}(z_{4},\bar{z}_{4})\rangle_{\lambda}
\\=&
\lambda\Big\{ h\bar{h}z_{13}^{2}\bar{z}_{13}^{2}{\cal I}_{2222}(z_{1},z_{3},\bar{z}_{1},\bar{z}_{3})+h\bar{h}z_{24}^{2}\bar{z}_{13}^{2}{\cal I}_{2222}(z_{2},z_{4},\bar{z}_{1},\bar{z}_{3})
\\&
+h\bar{h}z_{13}^{2}\bar{z}_{24}^{2}{\cal I}_{2222}(z_{1},z_{3},\bar{z}_{2},\bar{z}_{4})+h\bar{h}z_{24}^{2}\bar{z}_{24}^{2}{\cal I}_{2222}(z_{2},z_{4},\bar{z}_{2},\bar{z}_{4})
\\&
+\big(\bar{z}_{13}^{2}{\cal I}_{111122}(z_{1},z_{2},z_{3},z_{4},\bar{z}_{1},\bar{z}_{3})+\bar{z}_{24}^{2}{\cal I}_{111122}(z_{1},z_{2},z_{3},z_{4},\bar{z}_{2},\bar{z}_{4})\big)\bar{h}z_{23}z_{14}\frac{\eta\partial_{\eta}G(\eta,\bar{\eta})}{G(\eta,\bar{\eta})}
\\&
+\big(z_{13}^{2}{\cal I}_{221111}(z_{1},z_{3},\bar{z}_{1},\bar{z}_{3},\bar{z}_{2},\bar{z}_{4})+z_{24}^{2}{\cal I}_{221111}(z_{2},z_{4},\bar{z}_{2},\bar{z}_{4},\bar{z}_{1},\bar{z}_{3})\big)h\bar{z}_{23}\bar{z}_{14}\frac{\bar{\eta}\partial_{\bar{\eta}}G(\eta,\bar{\eta})}{G(\eta,\bar{\eta})}
\\&
+z_{23}z_{14}\bar{z}_{23}\bar{z}_{14}{\cal I}_{11111111}(z_{1},z_{2}z_{3},z_{4},\bar{z}_{1},\bar{z}_{3},\bar{z}_{2},\bar{z}_{4})\eta\bar{\eta}\frac{\partial_{\eta}\partial_{\bar{\eta}}G(\eta,\bar{\eta})}{G(\eta,\bar{\eta})}\Big\}
\\&
\langle{\cal O}^{\dagger}(z_{1},\bar{z}_{1}){\cal O}(z_{2},\bar{z}_{2}){\cal O}^{\dagger}(z_{3},\bar{z}_{3}){\cal O}(z_{4},\bar{z}_{4})\rangle.
\ea
\ee
We then could use the formula (\ref{eq:I2222z1-z2bz3bz4}), (\ref{eq:I221111}), (\ref{eq:I111122}) and (\ref{eq:I11111111}) in appendix to express this integral in terms of ${\cal I}_1$, ${\cal I}_2$ and ${\cal I}_3$. More precisely, the integral like ${\cal I}_{11}(z_1,\bar{z}_1)$ will also appear. The dimensional reduction parameter ${\epsilon}$ in ${\cal I}_1$ and  ${\cal I}_2$ is positive, while the parameter $\tilde{\epsilon}$ in ${\cal I}_3$ is negative. The integral ${\cal I}_{11}(z_1,\bar{z}_1)$ contains both $\epsilon$ and $\tilde{\epsilon}$. In our calculation, the contribution due to integral ${\cal I}_{11}(z_1,\bar{z}_1)$ will only replace the $\tilde{\epsilon}$ in ${\cal I}_3$ to ${\epsilon}$. So we can just ignore ${\cal I}_{11}(z_1,\bar{z}_1)$, and regard the parameter $\tilde{\epsilon}$ in paired ${\cal I}_3$ as positive\footnote{We would like to appreciate Yuan Sun to discuss with us on this issue.}. Then we use (\ref{eq:I1-dim-reg}), (\ref{eq:I2-regulated}) and (\ref{eq:I3-regulated}) to obtain the dimensional regulated result. Since the final result is quite complicated, we will not show the detail here.


\section{Entanglement entropy in the $T\bar{T}${-}deformed CFTs}
\label{sec:EE-TTbar}

The quantum entanglement of deformed CFTs have been investigated in \cite{Chen:2018eqk, Sun:2019ijq, Jeong:2019ylz}. However, these works are mainly focused on the R\'enyi entanglement entropy of vacuum states in deformed CFTs.
In this section, we first review the R\'enyi entanglement entropy of excited state in the un-deformed CFT and then consider their $T\bar{T}$-deformation.

Let us consider an excited state defined by acting a primary operator ${\cal O}_a$ on the vacuum state $\ket{0}$ in the two-dimensional CFT. We introduce the complex coordinate $(z,\bar{z})=(x+i\tau, x-i\tau)$, such that $x$ and $\tau$ are the Euclidean space and Euclidean time respectively. We insert the primary operator ${\cal O}_a$ at $x=-l<0$ and consider the real time-evolution from $\tau=0$ to $\tau=t$ with the Hamiltonian $H$ \cite{Nozaki:2014hna, He:2014mwa}. The corresponding density matrix is
\be
\ba
\label{eq:dm}
\rho(t)&={\cal N}e^{-iHt}e^{-\epsilon H}{\cal O}_{a}(-l)\ket{0}\bra{0}{\cal O}_{a}^{\dagger}(-l)e^{-\epsilon H}e^{iHt}\\&={\cal N}{\cal O}_{a}(w_{2},\bar{w}_{2})\ket{0}\bra{0}{\cal O}_{a}^{\dagger}(w_{1},\bar{w}_{1}),
\ea
\ee
where ${\cal N}$ is the normalization factor, $\epsilon$ is an ultraviolet regularization. Moreover, $w_1$ and $w_2$ are defined by
\be
\ba
w_{1}&=i(\epsilon-it)-l,\quad w_{2}=-i(\epsilon+it)-l\\
\bar{w}_{1}&=-i(\epsilon-it)-l,\quad\bar{w}_{2}=i(\epsilon+it)-l,
\ea
\ee
where $\epsilon \pm it$ are treated as the purely real numbers \cite{He:2014mwa}. In other words, we regarded $t$ as the pure imaginary number until the end of the calculation.

We then employ the replica method in the path integral formulation to compute the R\'enyi entanglement entropy. Let us choose the subsystem $A$ to be an interval $0\leq x\leq L$ at $\tau=0$. This leads to a $n$-sheet Riemann surface $\Sigma_n$ with $2n$-operators ${\cal O}_a$, i.e.
\be
\ba
S_A^{(n)}=\frac{1}{1-n}\log {\rm Tr}[\rho_A^n]=\frac{1}{1-n}\Big[\log\langle{\cal O}_{a}^{\dagger}(w_{1},\bar{w}_{1}){\cal O}_{a}(w_{2},\bar{w}_{2})\cdots{\cal O}_{a}^{\dagger}(w_{2n-1},\bar{w}_{2n-1}){\cal O}_{a}(w_{2n},\bar{w}_{2n})\rangle\Big].
\ea
\ee
 We are interested in the difference of ${S}_A^{(n)}$ between the excited state and the vacuum state:
 \be
 \ba
 \label{eq:n=2EXEE}
 \Delta S_{A}^{(n)}&=\frac{1}{1-n}\Big[\log\langle{\cal O}_{a}^{\dagger}(w_{1},\bar{w}_{1}){\cal O}_{a}(w_{2},\bar{w}_{2})\cdots{\cal O}_{a}^{\dagger}(w_{2n-1},\bar{w}_{2n-1}){\cal O}_{a}(w_{2n},\bar{w}_{2n})\rangle_{\Sigma_{n}}\\&\qquad-n\log\langle{\cal O}_{a}^{\dagger}(w_{1},\bar{w}_{1}){\cal O}_{a}(w_{2},\bar{w}_{2})\rangle_{\Sigma_{1}}\Big].
 \ea
 \ee
These quantities measure the effective quantum mechanical degrees of freedom of the operator\cite{Nozaki:2014hna, He:2014mwa} \footnote{We call the difference of the  $\Delta S_{A}^{(n)}$ as the excess of the R\'enyi  entanglement entropy.}.

 Let us consider the $n=2$ case. We apply the conformal transformation
 \be
 \label{eq:conf-map-wz}
 \frac{w}{w-L}=\big(\frac{z}{L}\big)^{2},
 \ee
such that $\Sigma_2$ is mapped to $\Sigma_1$. For this case, the coordinates $z_i$ are given by
\be
\ba
\label{eq:z-coordinate}
z_{1}&=-z_{3}=L \sqrt{\frac{l-t-i\epsilon}{l+L-t-i\epsilon}}\\
z_{2}&=-z_{4}=L \sqrt{\frac{l-t+i\epsilon}{l+L-t+i\epsilon}}.
\ea
\ee
On the $\Sigma_1$, the four-point function can be expressed as
\be
\label{eq:4pt-Sigma1}
\langle{\cal O}_{a}^{\dagger}(z_{1},\bar{z}_{1}){\cal O}_{a}(z_{2},\bar{z}_{2}){\cal O}_{a}^{\dagger}(z_{3},\bar{z}_{3}){\cal O}_{a}(z_{4},\bar{z}_{4})\rangle_{\Sigma_{1}}=|z_{13}z_{24}|^{-4h_{a}}G_{a}(\eta,\bar{\eta}),
\ee
where
\be
\eta=\frac{z_{12}z_{34}}{z_{13}z_{24}}, \quad \bar{\eta}=\frac{\bar{z}_{12}\bar{z}_{34}}{\bar{z}_{13}\bar{z}_{24}}.
\ee
We then apply the map (\ref{eq:conf-map-wz}) to express the four-point function on $\Sigma_2$:
\be
\ba
&\langle{\cal O}_{a}^{\dagger}(w_{1},\bar{w}_{1}){\cal O}_{a}(w_{2},\bar{w}_{2}){\cal O}_{a}^{\dagger}(w_{3},\bar{w}_{3}){\cal O}_{a}(w_{4},\bar{w}_{4})\rangle_{\Sigma_{2}}\\=&\prod_{i=1}^{4}|\frac{dw_{i}}{dz_{i}}|^{-2h_{a}}\langle{\cal O}_{a}^{\dagger}(z_{1},\bar{z}_{1}){\cal O}_{a}(z_{2},\bar{z}_{2}){\cal O}_{a}^{\dagger}(z_{3},\bar{z}_{3}){\cal O}_{a}(z_{4},\bar{z}_{4})\rangle_{\Sigma_{1}}
\ea
\ee

Here the R\'enyi entanglement entropy in $0<t<l$ or $t>L+l$ for ${\cal O}_a|0\rangle$ will be vanishing \cite{He:2014mwa}. When $l<t<L+l$, R\'enyi entanglement entropy will be the logarithm of the quantum dimension of the corresponding operator. Therefore, the excess of the R\'enyi entanglement entropy of locally excited states between the early time and late time is logarithmic quantum dimension of the local operator. In two-dimensional rational CFTs, the authors of \cite{He:2014mwa, Guo:2015uwa, Chen:2015usa, Guo:2018lqq, Apolo:2018oqv} obtain that the excess of R\'enyi entanglement entropy is a logarithmic quantum dimension of corresponding local operator which excites the space time \footnote{In two-dimensional  quantum gravity theory, e.g. Liouville field theory, one has to redefine the proper reference state \cite{He:2017lrg} to restore the causality structure and the excess of R\'enyi entanglement entropy will be the logarithm of the ratio of the fusion matrix elements between the exciting and reference states.}.

\subsection{$T\bar{T}${-}deformation}
In two-dimensional rational CFTs, we have investigated the excess of the R\'enyi entanglement entropy of locally excited states between the early time and late time is the logarithmic quantum dimension of the local operator, which have been proved to be universal. In this section, we consider the R\'enyi entanglement entropy of excited state in the $T\bar{T}${-}deformed CFT. Let us focus on the $T\bar{T}$-deformation of $\Delta S_A^{(2)}$ in (\ref{eq:n=2EXEE}). The vacuum state $\ket{0}$  will be deformed by the $T\bar{T}$ deformation. To take the deformation of the vacuum state, we also expand the vacuum state in deformation theory up to the first order. For the sake of simplicity, we start from the density matrix of the locally excited states and focus on the first order correction of R\'enyi entropy in $T\bar{T}${-}deformation by using the CFT perturbation theory. Therefore, we assume that the conformal transformation is still an approximated symmetry and make use of replica trick to obtain the R\'enyi entropy. We have to insert the $T\bar{T}${-}deformation operator in the $n$-sheeted
manifold. Since we only focus on the first order of the REE, combining the vacuum deformation and replica effects, it will give the $3n$ of the deformation on one sheet totally.
\be
\ba
&\quad -\big(\Delta S_{A,0}^{(2)}+\Delta S_{A,\lambda}^{(2)}\big)\\
&=\log\Big(\langle{\cal O}_{a}^{\dagger}(w_{1},\bar{w}_{1}){\cal O}_{a}(w_{2},\bar{w}_{2}){\cal O}_{a}^{\dagger}(w_{3},\bar{w}_{3}){\cal O}_{a}(w_{4},\bar{w}_{4})\rangle_{\Sigma_{2}}\\
&\qquad
+6\lambda\int d^{2}w\langle T\bar{T}(w,\bar{w}){\cal O}_{a}^{\dagger}(w_{1},\bar{w}_{1}){\cal O}_{a}(w_{2},\bar{w}_{2}){\cal O}_{a}^{\dagger}(w_{3},\bar{w}_{3}){\cal O}_{a}(w_{4},\bar{w}_{4})\rangle_{\Sigma_{2}}\Big)\\
&\quad -6\log\Big(\langle{\cal O}_{a}^{\dagger}(w_{1},\bar{w}_{1}){\cal O}_{a}(w_{2},\bar{w}_{2})\rangle_{\Sigma_{1}}+\lambda\int d^{2}w\langle T\bar{T}(w,\bar{w}){\cal O}_{a}^{\dagger}(w_{1},\bar{w}_{1}){\cal O}_{a}(w_{2},\bar{w}_{2})\rangle_{\Sigma_{1}}\Big).
\ea
\ee
To evaluate the correlator on the $\Sigma_2$, we use the conformal map (\ref{eq:conf-map-wz}) to map $w$ in $\Sigma_2$ to $z$ in $\Sigma$. Under the conformal map (\ref{eq:conf-map-wz}), the stress tensors transform as
\be
T(w)=(\frac{dz}{dw})^{2}T(z)+\frac{c}{12}\{z,w\},\quad\bar{T}(\bar{w})=(\frac{d\bar{z}}{d\bar{w}})^{2}\bar{T}(\bar{z})+\frac{c}{12}\{\bar{z},\bar{w}\},
\ee
where $\{z,w\}=\frac{z^{\prime\prime\prime}}{z^{\prime}}-\frac{3}{2}\frac{z^{\prime\prime2}}{z^{\prime2}}$ is the Schwarzian derivative. The $T\bar{T}$-operator thus transforms as
\be
\ba
T\bar{T}(w, \bar{w})&=\frac{(z^{2}-L^2)^{4}}{4L^{6}z^{2}}\frac{(\bar{z}^{2}-L^2)^{4}}{4L^{6}\bar{z}^{2}}\big(T(z)+\frac{c}{8z^2}\big)\big(\bar{T}(\bar{z})+\frac{c}{8\bar{z}^2}\big).
\ea
\ee
Using this transformation formula and expanding around $\lambda=0$, we find
\be
\ba
\label{defm-SA}
&\quad -\big(\Delta S_{A,0}^{(2)}+\Delta S_{A,\lambda}^{(2)}\big)\\
&=\log\Big(\prod_{i=1}^{4}|\frac{dw_{i}}{dz_{i}}|^{-2h_{a}}\frac{\langle{\cal O}_{a}^{\dagger}(z_{1},\bar{z}_{1}){\cal O}_{a}(z_{2},\bar{z}_{2}){\cal O}_{a}^{\dagger}(z_{3},\bar{z}_{3}){\cal O}_{a}(z_{4},\bar{z}_{4})\rangle_{\Sigma_{1}}}{\big(\langle{\cal O}_{a}^{\dagger}(w_{1},\bar{w}_{1}){\cal O}_{a}(w_{2},\bar{w}_{2})\rangle_{\Sigma_{1}}\big)^{2}}\Big)
\\
&+6\lambda\int d^{2}z\frac{(z^{2}-L^{2})^{2}(\bar{z}^{2}-L^{2})^{2}}{4L^{6}|z|^{2}}\frac{\langle\big(T(z)+\frac{c}{8z^2}\big)\big(\bar{T}(\bar{z})+\frac{c}{8\bar{z}^2}\big){\cal O}_{a}^{\dagger}(z_{1},\bar{z}_{1}){\cal O}_{a}(z_{2},\bar{z}_{2}){\cal O}_{a}^{\dagger}(z_{3},\bar{z}_{3}){\cal O}_{a}(z_{4},\bar{z}_{4})\rangle_{\Sigma_{1}}}{\langle{\cal O}_{a}^{\dagger}(z_{1},\bar{z}_{1}){\cal O}_{a}(z_{2},\bar{z}_{2}){\cal O}_{a}^{\dagger}(z_{3},\bar{z}_{3}){\cal O}_{a}(z_{4},\bar{z}_{4})\rangle_{\Sigma_{1}}}\\
&\quad-6\lambda\int d^{2}w\frac{\langle T\bar{T}(w,\bar{w}){\cal O}_{a}^{\dagger}(w_{1},\bar{w}_{1}){\cal O}_{a}(w_{2},\bar{w}_{2})\rangle_{\Sigma_{1}}}{\langle{\cal O}_{a}^{\dagger}(w_{1},\bar{w}_{1}){\cal O}_{a}(w_{2},\bar{w}_{2})\rangle_{\Sigma_{1}}}+{\cal O}(\lambda^2).
\ea
\ee

Let us focus on the large $c$ case.  The leading order is evaluated as
\be
 6\lambda\frac{c^{2}}{64}\int d^{2}z\frac{(z^{2}-L^{2})^{2} (\bar{z}^{2}-L^{2})^{2}}{4L^{6}|z|^{6}}\\
= 6\lambda\frac{c^{2}}{64}\frac{\pi}{2L^{6}}\int_{0}^{\infty}d\rho(\rho^{3}+4L^{4}\rho^{-1}+L^{8}\rho^{-5}).
\ee
To take the regularization, we introduce the cutoff by replacing $(0,\infty)$ to $(\frac{1}{\tilde{\Lambda}},\Lambda)$:
\be
\ba
\label{eq:leading-order}
 6\lambda\frac{c^{2}}{64} \int d^{2}z\frac{|z^{2}-L^{2}|^{4}}{4L^{6}|z|^{6}}&\xrightarrow{\mbox{cutoff}}
6\lambda\frac{c^{2}}{64}\frac{\pi}{L^{2}}(\frac{\Lambda^{4}+L^{8}\tilde{\Lambda}^{4}}{4}+4L^{4}\log\Lambda\tilde{\Lambda}).
\ea
\ee
In the leading order ${\cal O}(c^2)$, the R\'enyi entanglement entropy depends on the UV and IR cut off introduced by regularization. By using the Ward identity and the integrals in Appendix \ref{app:ExEE}, the order $c$ of eq. (\ref{defm-SA}) can be written as
\be
\ba
\label{eq:ExEE-next-order}
&\frac{c}{8}6\lambda\int d^{2}z\frac{(z^{2}-L^{2})^{2}(\bar{z}^{2}-L^{2})^{2}}{4L^{6}|z|^{2}}\frac{\langle\big(\frac{1}{z^{2}}\bar{T}(\bar{z})+\frac{1}{\bar{z}^{2}}T(z)\big){\cal O}_{a}^{\dagger}(z_{1},\bar{z}_{1}){\cal O}_{a}(z_{2},\bar{z}_{2}){\cal O}_{a}^{\dagger}(z_{3},\bar{z}_{3}){\cal O}_{a}(z_{4},\bar{z}_{4})\rangle_{\Sigma_{1}}}{\langle{\cal O}_{a}^{\dagger}(z_{1},\bar{z}_{1}){\cal O}_{a}(z_{2},\bar{z}_{2}){\cal O}_{a}^{\dagger}(z_{3},\bar{z}_{3}){\cal O}_{a}(z_{4},\bar{z}_{4})\rangle_{\Sigma_{1}}}\\
=&\frac{c}{8}6\lambda\Big\{-\frac{4\pi\bar{h}_a\log(\tilde{\Lambda}|z_{1}|)}{\bar{z}_{1}^{2}}-\frac{4\pi\bar{h}_a\bar{z}_{1}^{2}\log(\Lambda/|z_{1}|)}{L^{4}}+\frac{\bar{h}_a\pi}{L^{4}\bar{z}_{1}^{4}}(\bar{z}_{1}^{2}-L^{2})(\bar{z}_{1}^{6}L^{2}|z_{1}|^{-4}-|z_{1}|^{4})\\
&-\frac{4\pi\bar{h}_a\log(\tilde{\Lambda}|z_{2}|)}{\bar{z}_{2}^{2}}-\frac{4\pi\bar{h}_a\bar{z}_{2}^{2}\log(\Lambda/|z_{2}|)}{L^{4}}+\frac{\bar{h}_a\pi}{L^{4}\bar{z}_{2}^{4}}(\bar{z}_{2}^{2}-L^{2})(\bar{z}_{2}^{6}L^{2}|z_{2}|^{-4}-|z_{2}|^{4})\\&+2\pi\frac{\bar{z}_{23}\bar{z}_{14}}{4L^{6}}\frac{\bar{\eta}\partial_{\bar{\eta}}G_a(\eta,\bar{\eta})}{G_a(\eta,\bar{\eta})}\Big(\frac{4L^{4}\log(|z_{2}|/|z_{1}|)}{\bar{z}_{1}^{2}-\bar{z}_{2}^{2}}\\
&+\frac{2L^{6}\log(\tilde{\Lambda}|z_{1}|)}{\bar{z}_{1}^{2}\left(\bar{z}_{1}^{2}-\bar{z}_{2}^{2}\right)}-\frac{2L^{6}\log(\tilde{\Lambda|z_{2}|})}{\bar{z}_{2}^{2}\left(\bar{z}_{1}^{2}-\bar{z}_{2}^{2}\right)}-\frac{2\bar{z}_{1}^{2}L^{2}\log(\Lambda/|z_{1}|)}{\bar{z}_{1}^{2}-\bar{z}_{2}^{2}}+\frac{2\bar{z}_{2}^{2}L^{2}\log(\Lambda/|z_{2}|)}{\bar{z}_{1}^{2}-\bar{z}_{2}^{2}}\\
&-\frac{(\bar{z}_{2}^{2}-L^{2})^{2}}{\bar{z}_{2}^{4}(\bar{z}_{1}^{2}-\bar{z}_{2}^{2})}(-\frac{1}{4}|\bar{z}_{2}|^{4}+\frac{L^{4}}{4}|\bar{z}_{2}|^{-4}\bar{z}_{2}^{4})+\frac{(\bar{z}_{1}^{2}-L^{2})^{2}}{\bar{z}_{1}^{4}(\bar{z}_{1}^{2}-\bar{z}_{2}^{2})}(-\frac{1}{4}|\bar{z}_{1}|^{4}+\frac{L^{4}}{4}|\bar{z}_{1}|^{-4}\bar{z}_{1}^{4})\Big)\\
&-\frac{4\pi h_az_{1}^{2}\log(\Lambda/|z_{1}|)}{L^{4}}-\frac{4\pi h_a\log(|z_{1}|\tilde{\Lambda})}{z_{1}^{2}}+\frac{h_a\pi z_{1}^{2}}{L^{4}}\frac{1}{z_{1}^{6}}(L^{2}-z_{1}^{2})(-L^{2}z_{1}^{6}|z_{1}|^{-4}+|z_{1}|^{4})\\
&-\frac{4\pi h_az_{2}^{2}\log(\Lambda/|z_{2}|)}{L^{4}}-\frac{4\pi h_a\log(|z_{2}|\tilde{\Lambda})}{z_{1}^{2}}+\frac{h_a\pi}{L^{4}}\frac{1}{z_{2}^{4}}(L^{2}-z_{2}^{2})(-L^{2}z_{2}^{6}|z_{2}|^{-4}+|z_{2}|^{4})\\
&+\pi\frac{z_{23}z_{14}}{2L^{6}}\frac{\eta\partial_{\eta}G_a(\eta,\bar{\eta})}{G_a(\eta,\bar{\eta})}\Big(\frac{4L^{4}\log(|z_{2}|/|z_{1}|)}{z_{1}^{2}-z_{2}^{2}}\\
&+\frac{2L^{6}\log(\tilde{\Lambda}|z_{1}|)}{z_{1}^{2}\left(z_{1}^{2}-z_{2}^{2}\right)}-\frac{2L^{6}\log(\tilde{\Lambda}|z_{2}|)}{z_{2}^{2}\left(z_{1}^{2}-z_{2}^{2}\right)}-\frac{2L^{2}z_{1}^{2}\log(\Lambda/|z_{1}|)}{z_{1}^{2}-z_{2}^{2}}+\frac{2L^{2}z_{2}^{2}\log(\Lambda/|z_{2}|)}{z_{1}^{2}-z_{2}^{2}}\\
&+\frac{(L^{2}-z_{2}^{2})^{2}}{z_{2}^{4}(z_{1}^{2}-z_{2}^{2})}(\frac{1}{4}|z_{2}|^{4}-\frac{1}{4}L^{4}z_{2}^{4}|z_{2}|^{-4})-\frac{(L^{2}-z_{1}^{2})^{2}}{z_{1}^{4}(z_{1}^{2}-z_{2}^{2})}(\frac{1}{4}|z_{1}|^{4}-\frac{1}{4}L^{4}z_{1}^{4}|z_{1}|^{-4})\Big)\Big\}.
\ea
\ee

Then our task is to substitute the conformal block and evaluate the next order correction (\ref{eq:ExEE-next-order}). In generic CFTs, the function $G(\eta,\bar{\eta})$ in (\ref{eq:4pt-Sigma1}), can be expressed by using the conformal blocks \cite{Belavin:1984vu}
\be
G_a(\eta, \bar{\eta})=\sum_b (C^b_{aa})^2F_a(b|\eta)\bar{F}_a(b|\bar{\eta}),
\ee
where $b$ runs over all the primary operators.

When we take the early time in our setup $0< t<l$ or $t>L+l$, one finds the cross ratios
\be
\eta\sim \frac{L^2 \epsilon ^2}{4 (l-t)^2 (l+L-t)^2},\quad \bar{\eta}\sim \frac{L^2 \epsilon ^2}{4 (l+t)^2 (l+L+t)^2}.
\ee
In this limit, i.e. $(\eta, \bar{\eta})\to (0,0)$, the dominant contribution arise from the identity operator.  We thus get
\be
\label{eq:Ga-zzbar-00}
G_a(\eta, \bar{\eta})\sim |\eta|^{-4h_a},\quad (\eta, \bar{\eta})\to (0, 0).
\ee
Plugging this into (\ref{eq:ExEE-next-order}), we obtain the next order correction
\be
\ba
-\frac{c}{2}\lambda \frac{\pi  h L^2 t_e^2 (2 l+L)^2}{(l-t_e)^2 (l+t_e)^2 (l+L-t_e)^2 (l+L+t_e)^2}+{\cal O}(\epsilon^2)\quad \mbox{for}\quad 0<t_e<l\quad \mbox{or}\quad t_e>L+l.
\ea
\ee

As we take the late time $l< t< l+L$, the cross ratios behaves as \cite{He:2014mwa}
\be
\eta\sim 1-\frac{L^2 \epsilon ^2}{4 (l-t)^2 (l+L-t)^2},\quad \bar{\eta}\sim \frac{L^2 \epsilon ^2}{4 (l+t)^2 (l+L+t)^2}.
\ee
The conformal block at the limit $(\eta, \bar{\eta})\to (1,0)$ can be written as
\be
\label{eq:Ga-zzbar-10}
G_a(\eta, \bar{\eta})\sim F_{00}[a](1-\eta)^{-2 h_a} \bar{\eta}^{-2 h_a},
\ee
where $ F_{bc}[a]$ is a constant called as Fusion matrix \cite{Moore:1988uz, Moore:1988ss}. Substituting this into (\ref{eq:ExEE-next-order}), we find
\be
\ba
&-6\frac{c}{2}\lambda\frac{\pi  h L^2 t_l^2 (2 l+L)^2}{(l-t_l)^2 (l+t_l)^2 (l+L-t_l)^2 (l+L+t_l)^2}-6 \frac{c}{4}\lambda \epsilon \frac{\pi ^2 h L t_l  (2 l+L) \left(l (l+L)+t_l^2\right)}{(l-t_l)^2 (l+t_l)^2
   (l+L-t_l)^2 (l+L+t_l)^2} \\
  & \qquad\qquad\qquad \qquad   \qquad \qquad\qquad\qquad \qquad   \qquad \qquad\qquad\qquad \qquad  +{\cal O}(\epsilon^2), \qquad \mbox{for}\quad  l<t_l<L.
\ea
\ee
By using the Ward identity, at order ${\cal O}(c^0)$, the correction to the R\'enyi entanglement entropy of excited state can also be written in terms of integrals, which appear in a complicated form. We will not consider the correction in order ${\cal O}(c^0)$ in the present paper.

Together with the result at leading order (\ref{eq:leading-order}), we obtain the $T\bar{T}$-deformed $\Delta S_{A,\lambda}^{(2)}$ at large $c$ limit:
\be
\ba
-\Delta S_{A,\lambda}^{(2)}&=6 \lambda\frac{c^{2}}{64}\frac{\pi}{L^{2}}(\frac{\Lambda^{4}+L^{8}\tilde{\Lambda}^{4}}{4}+4L^{4}\log\Lambda\tilde{\Lambda})
-6\frac{c}{2}\lambda \frac{ \pi  h L^2 t^2 (2 l+L)^2}{(l-t)^2 (l+t)^2 (l+L-t)^2 (l+L+t)^2}\\
&\quad +{\cal O}(\epsilon, c^0, \lambda^2)\quad\quad\quad\quad \mbox{for}\qquad t>0.
\ea
\ee
The first term is associated with UV and IR cutoff. The second term is related to the nontrivial time dependence at the linear order of the central charge $c$. Comparing the R\'enyi entanglement entropy of excited state at the early time and late time, we find
\be
\ba
\label{eq:Ex-REE}
&\quad \Delta S_{A,\lambda}^{(2)}(t_l)-\Delta S_{A,\lambda}^{(2)}(t_e)\\
&=6\frac{c}{2}\lambda\Big(\frac{\pi  h L^2 t_l^2 (2 l+L)^2}{(l-t_l)^2 (l+t_l)^2 (l+L-t_l)^2 (l+L+t_l)^2}- \frac{\pi  h L^2 t_e^2 (2 l+L)^2}{(l-t_e)^2 (l+t_e)^2 (l+L-t_e)^2 (l+L+t_e)^2}\Big)\\
&\qquad\qquad+{\cal O}(\epsilon, \lambda, c^0),
\ea
\ee
where $t_e$ and $t_l$ label the early time and late time respectively. We thus find that at the leading order of $\lambda$ the excess of the R\'enyi entanglement entropy change dramatically in the order of ${\cal O}(c)$, which depends on the details of the CFT.


\section{OTOC in $T\bar{T}$-deformed CFTs}
The out of time order correlation function (OTOC) has been identified as a diagnostic of quantum chaos \cite{Shenker:2014cwa, Maldacena:2015waa, Roberts:2014ifa}. Remarkably, the field theory with Einstein gravity dual is proposed to exhibit the maximal Lyapunov exponent, which measures the growth rate of the OTOC. In this section, we investigate the OTOC between pairs of operators:
\be
\frac{\langle W(t)VW(t)V \rangle_\beta}{\langle W(t)W(t)\rangle_\beta \langle VV\rangle_\beta}
\ee
in the deformed CFTs to see whether the chaotic property is preserved or not after the $T\bar{T}-$deformation perturbatively. Since the OTOC can be broadly regarded as one of the quantities to capture the chaotic or integrable behavior, our study will shed light on the integrability/chaos after the $T\bar{T}$-deformation.

The thermal four-point correlator $\langle {\cal O}(x,t)\cdots\rangle_\beta$, $x,t$ are the coordinates of the spatially infinite thermal system, can be computed by the vacuum expectation values through the conformal transformation:
\be
\label{eq:thm-CFT}
\langle{\cal O}(x_{1},t_{1})\cdots\rangle_{\beta}=(\frac{2\pi z_{1}}{\beta})^{h}(\frac{2\pi\bar{z}_{1}}{\beta})^{\bar{h}}\langle{\cal O}(z_{1},\bar{z}_{1})\cdots\rangle,
\ee
where $z_i,\bar{z}_i$ are
\be
\label{eq:conf-map-OTOC}
z_{i}(x_{i},t_{i})=e^{\frac{2\pi}{\beta}(x_{i}+t_{i})},\quad\bar{z}_{i}(x_{i},t_{i})=e^{\frac{2\pi}{\beta}(x_{i}-t_{i})}.
\ee
We may deform the thermal system, i.e. two-dimensional CFT at finite temperature $1/\beta$, by inserting the $T\bar{T}$ operator The first order correction to the thermal correlator is
\be
\lambda\int d^{2}w\langle T\bar{T}(w,\bar{w}){\cal O}(w_{1},\bar{w}_{1})\cdots\rangle_{\beta}
\ee
where $w=x+t$ and $\bar{w}=x-t$. Taking account the transformation of the stress tensor transform under the conformal transformation, we find
\be
\ba
&\langle T\bar{T}(w,\bar{w}){\cal O}(w_{1},\bar{w}_{1})\cdots\rangle_{\beta}\\=&(\frac{2\pi z_{1}}{\beta})^{h}(\frac{2\pi\bar{z}_{1}}{\beta})^{\bar{h}}(\frac{2\pi z}{\beta})^{2}(\frac{2\pi\bar{z}}{\beta})^{2}\langle\big(T(z)-\frac{c}{24z^{2}}\big)\big(\bar{T}(\bar{z})-\frac{c}{24\bar{z}^{2}}\big){\cal O}(z_{1},\bar{z}_{1})\cdots\rangle
\ea
\ee

Under the $T\bar{T}$-deformation, the four-point function
\be
\frac{\langle W(w_{1},\bar{w}_{1})W(w_{2},\bar{w}_{2})V(w_{3},\bar{w}_{3})V(w_{4},\bar{w}_{4})\rangle_{\beta}}{\langle W(w_{1},\bar{w}_{1})W(w_{2},\bar{w}_{2})\rangle_{\beta}\langle V(w_{3},\bar{w}_{3})V(w_{4},\bar{w}_{4})\rangle_{\beta}}
\ee
is deformed to
\be
\ba
&\Big(\langle W(w_{1},\bar{w}_{1})W(w_{2},\bar{w}_{2})V(w_{3},\bar{w}_{3})V(w_{4},\bar{w}_{4})\rangle_{\beta}\\&\qquad
+\lambda\int d^{2}w_{a}\langle T\bar{T}(w_{a},\bar{w}_{a})W(w_{1},\bar{w}_{1})W(w_{2},\bar{w}_{2})V(w_{3},\bar{w}_{3})V(w_{4},\bar{w}_{4})\rangle_{\beta}\Big)\\
&\times\frac{1}{\Big(\langle W(w_{1},\bar{w}_{1})W(w_{2},\bar{w}_{2})\rangle_{\beta}+\lambda\int d^{2}w_{b}\langle T\bar{T}(w_{b},\bar{w}_{b})W(w_{1},\bar{w}_{1})W(w_{2},\bar{w}_{2})\rangle_{\beta}\Big)}\\
&\times\frac{1}{\Big(\langle V(w_{3},\bar{w}_{3})V(w_{4},\bar{w}_{4})\rangle_{\beta}+\lambda\int d^{2}w_{c}\langle T\bar{T}(w_{c},\bar{w}_{c})V(w_{3},\bar{w}_{3})V(w_{4},\bar{w}_{4})\rangle_{\beta}\Big)}.
\ea
\ee
Expanding around $\lambda=0$ and performing the coordinate transformation (\ref{eq:conf-map-OTOC}), we obtain
\be
\ba
&\frac{\langle W(w_{1},\bar{w}_{1})W(w_{2},\bar{w}_{2})V(w_{3},\bar{w}_{3})V(w_{4},\bar{w}_{4})\rangle_{\beta}}{\langle W(w_{1},\bar{w}_{1})W(w_{2},\bar{w}_{2})\rangle_{\beta}\langle V(w_{3},\bar{w}_{3})V(w_{4},\bar{w}_{4})\rangle_{\beta}}\\
&\times\Big(1-\lambda(\frac{2\pi}{\beta})^{2}\int d^{2}z_{b}|z_{b}|^{2}\frac{\langle\big(T(z_{b})-\frac{c}{24z^{2}}\big)\big(\bar{T}(\bar{z}_{b})-\frac{c}{24\bar{z}^{2}}\big)W(z_{1},\bar{z}_{1})W(z_{2},\bar{z}_{2})\rangle}{\langle W(z_{1},\bar{z}_{1})W(z_{2},\bar{z}_{2})\rangle}\\
&-\lambda(\frac{2\pi}{\beta})^{2}\int d^{2}z_{c}|z_{c}|^{2}\frac{\langle\big(T(z_{c})-\frac{c}{24z^{2}}\big)\big(\bar{T}(\bar{z}_{c})-\frac{c}{24\bar{z}^{2}}\big)V(z_{3},\bar{z}_{3})V(z_{4},\bar{z}_{4})\rangle}{\langle V(z_{3},\bar{z}_{3})V(z_{4},\bar{z}_{4})\rangle}\\
&+\lambda(\frac{2\pi}{\beta})^{2}\int d^{2}z_{a}|z_{a}|^{2}\frac{\langle\big(T(z_{a})-\frac{c}{24z^{2}}\big)\big(\bar{T}(\bar{z}_{a})-\frac{c}{24\bar{z}^{2}}\big)W(z_{1},\bar{z}_{1})W(z_{2},\bar{z}_{2})V(z_{3},\bar{z}_{3})V(z_{4},\bar{z}_{4})\rangle}{\langle W(z_{1},\bar{z}_{1})W(z_{2},\bar{z}_{2})V(z_{3},\bar{z}_{3})V(z_{4},\bar{z}_{4})\rangle}+{\cal O}(\lambda^{2})\Big)\\
\ea
\ee
The term of order ${\cal O}(c^2)$ thus can be written as
\be
\ba
\label{eq:OTOC-Oc2}
&\lambda \frac{c^{2}}{24^{2}} (\frac{2\pi}{\beta})^{2} \big(\int d^{2}z_{a}|z_{a}|^{-2} -\int d^{2}z_{b}|z_{b}|^{-2}-\int d^{2}z_{c}|z_{c}|^{-2}\big)\\
\xrightarrow{{\rm cutoff}}
&-\lambda(\frac{2\pi}{\beta})^{2}\frac{c^{2}}{24^{2}}2\pi\int_{\frac{1}{\tilde{\Lambda}}}^{\Lambda}d\rho\frac{1}{\rho}=-\lambda(\frac{2\pi}{\beta})^{2}\frac{c^{2}}{24^{2}}2\pi\log(\Lambda\tilde{\Lambda}).
\ea
\ee
Note that this divergence only depends on the cutoff. Since no dynamics appear, this is not interested for us.

We then consider the order ${\cal O}(c)$
\be
\ba
\label{eq:OTOC-next-order}
&\lambda\frac{c}{24}(\frac{2\pi}{\beta})^{2}\Big\{
-\int d^{2}z_{a}|z_{a}|^{2}\frac{1}{\bar{z}_{a}^{2}}\frac{\langle T(z_{a})W(z_{1},\bar{z}_{1})W(z_{2},\bar{z}_{2})V(z_{3},\bar{z}_{3})V(z_{4},\bar{z}_{4})\rangle}{\langle W(z_{1},\bar{z}_{1})W(z_{2},\bar{z}_{2})V(z_{3},\bar{z}_{3})V(z_{4},\bar{z}_{4})\rangle}\\
&\qquad\qquad\quad -\lambda\int d^{2}z_{a}|z_{a}|^{2}\frac{1}{z_{a}^{2}}\frac{\langle\bar{T}(\bar{z}_{a})W(z_{1},\bar{z}_{1})W(z_{2},\bar{z}_{2})V(z_{3},\bar{z}_{3})V(z_{4},\bar{z}_{4})\rangle}{\langle W(z_{1},\bar{z}_{1})W(z_{2},\bar{z}_{2})V(z_{3},\bar{z}_{3})V(z_{4},\bar{z}_{4})\rangle}\\
&\qquad+ \int d^{2}z_{b}|z_{b}|^{2}\frac{1}{\bar{z}_{b}^{2}}\frac{\langle T(z_{b})W(z_{1},\bar{z}_{1})W(z_{2},\bar{z}_{2})\rangle}{\langle W(z_{1},\bar{z}_{1})W(z_{2},\bar{z}_{2})\rangle}+\int d^{2}z_{b}|z_{b}|^{2}\frac{1}{z_{b}^{2}}\frac{\langle\bar{T}(\bar{z}_{b})W(z_{1},\bar{z}_{1})W(z_{2},\bar{z}_{2})\rangle}{\langle W(z_{1},\bar{z}_{1})W(z_{2},\bar{z}_{2})\rangle}\\
&\qquad +\int d^{2}z_{c}|z_{c}|^{2}\frac{1}{\bar{z}_{c}^{2}}\frac{\langle\big(T(z_{c})\big)V(z_{3},\bar{z}_{3})V(z_{4},\bar{z}_{4})\rangle}{\langle V(z_{3},\bar{z}_{3})V(z_{4},\bar{z}_{4})\rangle}+\int d^{2}z_{c}|z_{c}|^{2}\frac{1}{z_{c}^{2}}\frac{\langle\bar{T}(\bar{z}_{c})V(z_{3},\bar{z}_{3})V(z_{4},\bar{z}_{4})\rangle}{\langle V(z_{3},\bar{z}_{3})V(z_{4},\bar{z}_{4})\rangle}\Big\}.
\ea
\ee
Note that the four-point function in un-deformed CFT is given by
\be
\ba
\langle W(z_{1},\bar{z}_{1})W(z_{2},\bar{z}_{2})V(z_{3},\bar{z}_{3})V(z_{4},\bar{z}_{4})\rangle=\frac{1}{z_{12}^{2h_{w}}z_{34}^{2h_{v}}}\frac{1}{\bar{z}_{12}^{2\bar{h}_{w}}\bar{z}_{34}^{2\bar{h}_{v}}}G(\eta,\bar{\eta}).
\ea
\ee
The two-point function in un-deformed CFT behaves as
\be
\ba
\langle W(z_{1},\bar{z}_{1})W(z_{2},\bar{z}_{2})\rangle=\frac{1}{z_{12}^{2h_{w}}\bar{z}_{12}^{2\bar{h}_{w}}}.
\ea
\ee
The two-point functions for operator $V$ also has the similar construction. Using the Ward identity and the integrals in Appendix \ref{app:OTOC}, we evaluate the next order correction (\ref{eq:OTOC-next-order}) as
\be
\ba
\label{eq:OTOC-Oc-G}
&\lambda\frac{c}{24}(\frac{2\pi}{\beta})^{2}\Big\{\\
&-2\pi\frac{\eta\partial_{\eta}G(\eta,\eta)}{G(\eta,\eta)}z_{14}z_{23}\Big(\frac{z_{1}}{z_{12}z_{13}z_{14}}\log\frac{1}{|z_{1}|}+\frac{z_{3}}{z_{13}z_{23}z_{34}}\log\frac{1}{|z_{3}|}-\frac{z_{4}}{z_{14}z_{24}z_{34}}\log\frac{1}{|z_{4}|}-\frac{z_{2}}{z_{12}z_{23}z_{24}}\log\frac{1}{|z_{2}|}\Big)
\\&
+2\pi\frac{\bar{\eta}\partial_{\bar{\eta}}G(\eta,\eta)}{G(\eta,\eta)}\bar{z}_{14}\bar{z}_{23}\Big(\frac{\bar{z}_{1}}{\bar{z}_{12}\bar{z}_{13}\bar{z}_{14}}\log|z_{1}|-\frac{\bar{z}_{2}}{\bar{z}_{12}\bar{z}_{23}\bar{z}_{24}}\log|z_{2}|+\frac{\bar{z}_{3}}{\bar{z}_{13}\bar{z}_{23}\bar{z}_{34}}\log|z_{3}|-\frac{\bar{z}_{4}}{\bar{z}_{14}\bar{z}_{24}\bar{z}_{34}}\log|z_{4}|\Big)\Big\}.
\ea
\ee

Then our task is to evaluate (\ref{eq:OTOC-Oc-G}).  In the two-dimensional CFT, $G(\eta,\bar{\eta})$ can expand in terms of global conformal blocks\cite{Dolan:2000ut}:
\be
\ba
G(\eta,\bar{\eta})&=\sum_{h,\bar{h}}p(h,\bar{h})\eta^{h}\bar{\eta}^{\bar{h}}F(h,h,2h,\eta)F(\bar{h},\bar{h},2\bar{h},\bar{\eta}),
\ea
\ee
where $F$ is the Gauss hypergeometric function. The summation is over the global $SL(2)$ primary operator. The coefficient $p$ is related to operator expansion coefficient $p(h,\bar{h})=\lambda_{WW{\cal O}_{h,\bar{h}}}\lambda_{VV{\cal O}_{h,\bar{h}}}$.

For the two-dimensional CFT corresponding to the Einstein gravity theory, all the desired propagation can be expressed by using the identity operator, and the conformal block can be replaced by
\be
G(\eta,\bar{\eta}) \to {\cal F}(\eta) {\cal F}(\bar{\eta})
\ee
where ${\cal F}$ is the Virasoro conformal block whose dimension is zero in the intermediate channel. Here we will use a slight different notation compared with previous section. The function ${\cal F}$ is not known in generic cases. However, at large $c$ with small $h_w/c$ fixed and large $h_v$ fixed, the formula reads \cite{Fitzpatrick:2014vua}
\be
{\cal F}(\eta)  \sim \Big(\frac{\eta(1-\eta)^{-6h_{w}/c}}{1-(1-\eta)^{1-12h_{w}/c}}\Big)^{2h_{v}},
\ee
where the function has a branch cut at $\eta=1$. For the contour around $\eta=1$ and small $\eta$, one finds
\be
\label{eq:CB-F}
{\cal F}(\eta)\sim \Big(\frac{1}{1-\frac{24\pi ih_{w}}{c\eta}}\Big)^{2h_{v}}.
\ee
Since the path of $\bar{\eta}$ does not cross the brach cut at $\bar{\eta}=1$, one find $\bar{\cal F}(\bar{\eta})=1$ at small $\bar{\eta}$.

To apply the $T\bar{T}${-}deformed correlation function to the OTOC, we follow the steps in \cite{Roberts:2014ifa, Perlmutter:2016pkf} to evaluate the OTOC by using the analytic of the Euclideans of the four-point function by writing
\be
\ba
\label{eq:OTOC-z}
z_{1}&=e^{\frac{2\pi}{\beta}i\epsilon_{1}},\quad\bar{z}_{1}=e^{-\frac{2\pi}{\beta}i\epsilon_{1}}\\
z_{2}&=e^{\frac{2\pi}{\beta}i\epsilon_{2}},\quad\bar{z}_{2}=e^{-\frac{2\pi}{\beta}i\epsilon_{2}}\\
z_{3}&=e^{\frac{2\pi}{\beta}(t+i\epsilon_{3}-x)},\quad\bar{z}_{3}=e^{\frac{2\pi}{\beta}(-t-i\epsilon_{3}-x)}\\
z_{4}&=e^{\frac{2\pi}{\beta}(t+i\epsilon_{4}-x)},\quad\bar{z}_{4}=e^{\frac{2\pi}{\beta}(-t-i\epsilon_{4}-x)}
\ea
\ee
as the function of the continuation parameter $t$. Substituting the coordinates (\ref{eq:OTOC-z}) and (\ref{eq:CB-F}) to (\ref{eq:OTOC-Oc-G}), we find
\be
\ba
\label{eq:OTOC-Oc}
&\lambda\frac{c}{24}(\frac{2\pi}{\beta})^{2}\Big\{\\
&\int d^{2}z_{b}|z_{b}|^{2}\frac{1}{\bar{z}_{b}^{2}}\frac{\langle T(z_{b})W(z_{1},\bar{z}_{1})W(z_{2},\bar{z}_{2})\rangle}{\langle W(z_{1},\bar{z}_{1})W(z_{2},\bar{z}_{2})\rangle}+\int d^{2}z_{b}|z_{b}|^{2}\frac{1}{z_{b}^{2}}\frac{\langle\bar{T}(\bar{z}_{b})W(z_{1},\bar{z}_{1})W(z_{2},\bar{z}_{2})\rangle}{\langle W(z_{1},\bar{z}_{1})W(z_{2},\bar{z}_{2})\rangle}\\
&+\int d^{2}z_{c}|z_{c}|^{2}\frac{1}{\bar{z}_{c}^{2}}\frac{\langle\big(T(z_{c})\big)V(z_{3},\bar{z}_{3})V(z_{4},\bar{z}_{4})\rangle}{\langle V(z_{3},\bar{z}_{3})V(z_{4},\bar{z}_{4})\rangle}+\int d^{2}z_{c}|z_{c}|^{2}\frac{1}{z_{c}^{2}}\frac{\langle\bar{T}(\bar{z}_{c})V(z_{3},\bar{z}_{3})V(z_{4},\bar{z}_{4})\rangle}{\langle V(z_{3},\bar{z}_{3})V(z_{4},\bar{z}_{4})\rangle}\\
&-\int d^{2}z_{a}|z_{a}|^{2}\frac{1}{\bar{z}_{a}^{2}}\frac{\langle T(z_{a})W(z_{1},\bar{z}_{1})W(z_{2},\bar{z}_{2})V(z_{3},\bar{z}_{3})V(z_{4},\bar{z}_{4})\rangle}{\langle W(z_{1},\bar{z}_{1})W(z_{2},\bar{z}_{2})V(z_{3},\bar{z}_{3})V(z_{4},\bar{z}_{4})\rangle}\\
&-\lambda\int d^{2}z_{a}|z_{a}|^{2}\frac{1}{z_{a}^{2}}\frac{\langle\bar{T}(\bar{z}_{a})W(z_{1},\bar{z}_{1})W(z_{2},\bar{z}_{2})V(z_{3},\bar{z}_{3})V(z_{4},\bar{z}_{4})\rangle}{\langle W(z_{1},\bar{z}_{1})W(z_{2},\bar{z}_{2})V(z_{3},\bar{z}_{3})V(z_{4},\bar{z}_{4})\rangle}\Big\}\\
=& c\lambda h_{w}h_{v}\frac{8\pi^{5}}{\beta^{3}}xe^{\frac{4\pi x}{\beta}}\frac{e^{\frac{4\pi\left(t-x+i\epsilon_{3}\right)}{\beta}}-1}{6\pi h_{w}\big(e^{\frac{2\pi x}{\beta}}-e^{\frac{2\pi(t+i\epsilon_{3})}{\beta}}\big){}^{2}-ice^{\frac{2\pi(t+x+i\epsilon_{3})}{\beta}}},
\ea
\ee
where we have located the operators in pairs: $\epsilon_2=\epsilon_1+\beta/2$ and $\epsilon_4=\epsilon_3+\beta/2$, and set $\epsilon_1=0$ without loss generality \cite{Perlmutter:2016pkf}.

Let us now consider the order ${\cal O}(c^0)$ correction. By using the Ward identity, we find
\be
\ba
\label{eq:OTOC-O1}
&\lambda(\frac{2\pi}{\beta})^{2}\int d^{2}z|z|^{2}\Big\{\frac{\langle T(z)\bar{T}(\bar{z})W(z_{1},\bar{z}_{1})W(z_{2},\bar{z}_{2})V(z_{3},\bar{z}_{3})V(z_{4},\bar{z}_{4})\rangle}{\langle W(z_{1},\bar{z}_{1})W(z_{2},\bar{z}_{2})V(z_{3},\bar{z}_{3})V(z_{4},\bar{z}_{4})}\\
&-\frac{\langle T(z)\bar{T}(\bar{z})W(z_{1},\bar{z}_{1})W(z_{2},\bar{z}_{2})\rangle}{\langle W(z_{1},\bar{z}_{1})W(z_{2},\bar{z}_{2})\rangle}-\frac{\langle T(z)\bar{T}(\bar{z})V(z_{3},\bar{z}_{3})V(z_{4},\bar{z}_{4})\rangle}{\langle V(z_{3},\bar{z}_{3})V(z_{4},\bar{z}_{4})}\Big\}\\
=&\lambda(\frac{2\pi}{\beta})^2\Big\{12h_{v}^{2}h_{w}\frac{e^{-\frac{4\pi(t+i\epsilon_{3})}{\beta}}\big(e^{\frac{4\pi(t+i\epsilon_{3})}{\beta}}+e^{\frac{8\pi(t+i\epsilon_{3})}{\beta}}+e^{\frac{4\pi(t+2x+i\epsilon_{3})}{\beta}}-2e^{\frac{4\pi x}{\beta}}-1\big)}{(e^{\frac{4\pi x}{\beta}}+1)\big(-ic+12\pi h_{w}\cosh\left(\frac{2\pi\left(t-x+i\epsilon_{3}\right)}{\beta}\right)-12\pi h_{w}\big)\sinh\big(\frac{2\pi(t+x+i\epsilon_{3})}{\beta}\big)}\\
&+12h_{v}h_{w}^{2}\frac{e^{\frac{4\pi(t+i\epsilon_{3})}{\beta}}+e^{\frac{4\pi(t+2x+i\epsilon_{3})}{\beta}}-3e^{\frac{4\pi(2t+x+2i\epsilon_{3})}{\beta}}+e^{\frac{4\pi x}{\beta}}}{\big(-1+e^{\frac{4\pi(t+x+i\epsilon_{3})}{\beta}}\big)\Big(ice^{\frac{2\pi\left(t+x+i\epsilon_{3}\right)}{\beta}}-6\pi h_{w}\big(e^{\frac{2\pi x}{\beta}}-e^{\frac{2\pi(t+i\epsilon_{3})}{\beta}}\big){}^{2}\Big)}\\
&-\frac{24h_{v}h_{w}}{\epsilon^{2}}\frac{\big(-e^{\frac{4\pi x}{\beta}}+e^{\frac{4\pi(t+i\epsilon_{3})}{\beta}}\big)\big(h_{v}-h_{w}e^{\frac{4\pi x}{\beta}}\big)}{6\pi h_{w}e^{\frac{4\pi x}{\beta}}\big(e^{\frac{2\pi x}{\beta}}-e^{\frac{2\pi(t+i\epsilon_{3})}{\beta}}\big){}^{2}-ice^{\frac{2\pi(t+3x+i\epsilon_{3})}{\beta}}}\Big\}.
\ea
\ee
where $\epsilon$ is the cutoff denoted by $|z_i|^2=z_i \bar{z}_i+\epsilon^2$. Taking together with (\ref{eq:OTOC-Oc2}), (\ref{eq:OTOC-Oc}) and (\ref{eq:OTOC-O1}), we find the $T\bar{T}$-deformed OTOC at late time behaves  as
\be
\ba
&\frac{\langle W(w_{1},\bar{w}_{1})W(w_{2},\bar{w}_{2})V(w_{3},\bar{w}_{3})V(w_{4},\bar{w}_{4})\rangle_{\beta}}{\langle W(w_{1},\bar{w}_{1})W(w_{2},\bar{w}_{2})\rangle_{\beta}\langle V(w_{3},\bar{w}_{3})V(w_{4},\bar{w}_{4})\rangle_{\beta}}\\
\xrightarrow{T\bar{T}}& \frac{\langle W(w_{1},\bar{w}_{1})W(w_{2},\bar{w}_{2})V(w_{3},\bar{w}_{3})V(w_{4},\bar{w}_{4})\rangle_{\beta}}{\langle W(w_{1},\bar{w}_{1})W(w_{2},\bar{w}_{2})\rangle_{\beta}\langle V(w_{3},\bar{w}_{3})V(w_{4},\bar{w}_{4})\rangle_{\beta}}
\Big\{1-\lambda C_1(x)+\lambda C_2(x) e^{-\frac{2\pi}{\beta}t}+\cdots\Big\},
\ea
\ee
where $C_1(x)$ and $C_2(x)$ are the terms independent of $t$. Therefore, the Lyapunov exponent, which measures the time growth rate, is not affected. Further, the choices of the sign of $\lambda$ do not affect the late time behavior $e^{-\frac{2\pi}{\beta}t}$ in the above equation.

We thus expect the $T\bar{T}$-deformation does not affect the maximal chaos found by OTOC up to the perturbation first order of the deformation \footnote{A similar result in AdS$_2$ and Schwarzian theory has been found recently in \cite{Gross:2019ach}}. Moreover, since the bound of the Lyapunov exponent found in OTOC is un-affect, the gravity dual of the $T\bar{T}$-deformed holographic CFT is expected to  saturate the bound of the chaos. Here we focus on the late time behavior of the OTOC in the $T\bar{T}$ deformed large central charge CFT which is expected to have holographic dual. Although we have not investigated the integral model directly, it is also natural to expect that the $T\bar{T}$-deformed integrable model is still integrable up to the perturbation first order of the deformation.


\section{$J\bar{T}${-}deformation}
\subsection{Correlation functions in $J\bar{T}$-deformed CFTs}
It is also interesting to consider the deformation of $J\bar{T}$, {which is defined by adding an operator constructed from a chiral $U(1)$ current $J$ and stress tensor $\bar{T}$ in the action
\be
\frac{d S}{d\lambda}=\int d^2 z(J\bar{T})_\lambda.
\ee} {In a similar way as in $T\bar{T}$-deformation, we regard the deformation as a perturbative theory, in which case the action can be written as
\be
S(\lambda)=S(\lambda=0)+\lambda\int d^2z\sqrt{g}J\bar{T}+{\cal O}(\lambda^2),
\ee
where we denoted $(J\bar{T})_{\lambda=0}=J\bar{T}$. The first order correction to the correlation function is
\be
\ba
\langle {\cal O}_{1}(z_{1},\bar{z}_{1})\cdots{\cal O}_{n}(z_{n},\bar{z}_{n}) \rangle_\lambda=\lambda \int d^2 z \langle J\bar{T}(z, \bar{z}){\cal O}_{1}(z_{1},\bar{z}_{1})\cdots{\cal O}_{n}(z_{n},\bar{z}_{n})\rangle.
\ea
\ee}
By using the Ward identity, this correction becomes
\be
\ba
\langle {\cal O}_{1}(z_{1},\bar{z}_{1})\cdots{\cal O}_{n}(z_{n},\bar{z}_{n})\rangle_\lambda&= \lambda \int d^2 z \Big(\sum_{i=1}^{n}\frac{q_{i}}{z-z_{i}}\Big)\Big(\sum_{i=1}^{n}\big(\frac{\bar{h}_{i}}{(\bar{z}-\bar{z}_{i})^{2}}+\frac{\partial_{\bar{z}_{i}}}{\bar{z}-\bar{z}_{i}}\big)\Big)\langle{\cal O}_{1}(z_{1},\bar{z}_{1})\cdots{\cal O}_{n}(z_{n},\bar{z}_{n})\rangle,
\ea
\ee
where ${\cal O}_i$ is the primary operator with dimension $(h,\bar{h})$ and charge $q$. Therefore, the first order correction of two-point correlator due to $J\bar{T}${-}deformation is
\be
\langle {\cal O}(z_{1},\bar{z}_{1}){\cal O}^{\dagger}(z_{2},\bar{z}_{2})\rangle_\lambda=\lambda\int d^{2}z\Big(\frac{q_{1}}{z-z_{1}}+\frac{q_{2}}{z-z_{2}}\Big)\frac{\bar{h}\bar{z}_{12}^{2}}{(\bar{z}-\bar{z}_{1})^{2}(\bar{z}-\bar{z}_{2})^{2}}\langle{\cal O}(z_{1},\bar{z}_{1}){\cal O}^{\dagger}(z_{2},\bar{z}_{2})\rangle.
\ee
By using the integral (\ref{eq:I122z1bz1bz3}), we can express this correction in terms of ${\cal I}_3$, which is evaluated by using the dimensional regularization.

It is easy to find the first order correction to the four-point function is
\be
\ba
&\langle {\cal O}^{\dagger}(z_{1},\bar{z}_{1}){\cal O}(z_{2},\bar{z}_{2}){\cal O}^{\dagger}(z_{3},\bar{z}_{3}){\cal O}(z_{4},\bar{z}_{4})\rangle_\lambda\\
=&\lambda\int d^{2}z\Big(\sum_{i=1}^{4}\frac{q_{i}}{z-z_{i}}\Big)\Big(\frac{\bar{h}_{a}\bar{z}_{13}^{2}}{(\bar{z}-\bar{z}_{1})^{2}(\bar{z}-\bar{z}_{3})^{2}}+\frac{\bar{h}_{a}\bar{z}_{24}^{2}}{(\bar{z}-\bar{z}_{2})^{2}(\bar{z}-\bar{z}_{4})^{2}}+\frac{\bar{z}_{23}\bar{z}_{14}}{\prod_{j=1}^{4}(\bar{z}-\bar{z}_{j})}\frac{\bar{\eta}\partial_{\bar{\eta}}G(\eta,\bar{\eta})}{G(\eta,\bar{\eta})}\Big)\\
&\langle{\cal O}^{\dagger}(z_{1},\bar{z}_{1}){\cal O}(z_{2},\bar{z}_{2}){\cal O}^{\dagger}(z_{3},\bar{z}_{3}){\cal O}(z_{4},\bar{z}_{4})\rangle.
\ea
\ee
By using the integrals (\ref{eq:integrals}), we express this correction as
\be
\ba
&\langle {\cal O}^{\dagger}(z_{1},\bar{z}_{1}){\cal O}(z_{2},\bar{z}_{2}){\cal O}^{\dagger}(z_{3},\bar{z}_{3}){\cal O}(z_{4},\bar{z}_{4})\rangle_\lambda\\
=&\lambda\Big(\bar{h}_{a}\bar{z}_{13}^{2}\big(q_{1}{\cal I}_{122}(z_{1},\bar{z}_{1},\bar{z}_{3})+q_{3}{\cal I}_{122}(z_{3},\bar{z}_{1},\bar{z}_{3})+q_{2}{\cal I}_{122}(z_{2},\bar{z}_{1},\bar{z}_{3})+q_{4}{\cal I}_{122}(z_{4},\bar{z}_{1},\bar{z}_{3})\big)\\&+\bar{h}_{a}\bar{z}_{24}^{2}\big(q_{1}{\cal I}_{122}(z_{1},\bar{z}_{2},\bar{z}_{4})+q_{2}{\cal I}_{122}(z_{2},\bar{z}_{2},\bar{z}_{4})+q_{3}{\cal I}_{122}(z_{3},\bar{z}_{2},\bar{z}_{4})+q_{4}{\cal I}_{122}(z_{4},\bar{z}_{2},\bar{z}_{4})\big)\\&+\big(\sum_{i=1}^{4}q_{i}{\cal I}_{11111}(z_{i},\bar{z}_{1},\bar{z}_{2},\bar{z}_{3},\bar{z}_{4})\Big)\bar{z}_{23}\bar{z}_{14}\frac{\bar{\eta}\partial_{\bar{\eta}}G(\eta,\bar{\eta})}{G(\eta,\bar{\eta})}\langle{\cal O}^{\dagger}(z_{1},\bar{z}_{1}){\cal O}(z_{2},\bar{z}_{2}){\cal O}^{\dagger}(z_{3},\bar{z}_{3}){\cal O}(z_{4},\bar{z}_{4})\rangle.
\ea
\ee
Using the formulas (\ref{eq:I4-122}), (\ref{eq:I122z1bz1bz3}) and (\ref{eq:I11111}), we could express this integral in terms of ${\cal I}_3$. Since the final result is quite complicated, we will not show the details here.

\subsection{Entanglement entropy in $J\bar{T}$-deformed CFTs}
In this section, we consider the R\'enyi  entanglement entropy of excited state in the $J\bar{T}${-}deformed CFT. In the parallel with the $T\bar{T}$-deformation, we consider  the $J\bar{T}$-deformation of $\Delta S_A^{(2)}$ in (\ref{eq:n=2EXEE}).
\be
\ba
-(\Delta S_{A,0}^{(2)}+\Delta S_{A,\lambda}^{(2)})&=\log\Big(\langle{\cal O}_{a}^{\dagger}(w_{1},\bar{w}_{1}){\cal O}_{a}(w_{2},\bar{w}_{2}){\cal O}_{a}^{\dagger}(w_{3},\bar{w}_{3}){\cal O}_{a}(w_{4},\bar{w}_{4})\rangle_{\Sigma_{2}}\\
&\quad+6\lambda\int d^{2}w\langle J\bar{T}(w,\bar{w}){\cal O}_{a}^{\dagger}(w_{1},\bar{w}_{1}){\cal O}_{a}(w_{2},\bar{w}_{2}){\cal O}_{a}^{\dagger}(w_{3},\bar{w}_{3}){\cal O}_{a}(w_{4},\bar{w}_{4})\rangle_{\Sigma_{2}}\Big)\\
&\quad-6\log\Big(\langle{\cal O}_{a}^{\dagger}(w_{1},\bar{w}_{1}){\cal O}_{a}(w_{2},\bar{w}_{2})\rangle_{\Sigma_{1}}+\lambda\int d^{2}w\langle J\bar{T}(w,\bar{w}){\cal O}_{a}^{\dagger}(w_{1},\bar{w}_{1}){\cal O}_{a}(w_{2},\bar{w}_{2})\rangle_{\Sigma_{1}}\Big).
\ea
\ee
To evaluate the correlator on the $\Sigma_2$, we use the conformal map (\ref{eq:conf-map-wz}) to map $w$ in $\Sigma_2$ to $z$ in $\Sigma_1$. Under the conformal map, the current and the stress tensors transform as
\be
\ba
J\bar{T}(w,\bar{w})&=\frac{dz}{dw}J(z)\frac{(L^{2}-\bar{z}^{2})^{4}}{4L^{6}\bar{z}^{2}}\big(\bar{T}(\bar{z})+\frac{c}{8\bar{z}^{2}}\big).
\ea
\ee
Expanding around $\lambda=0$, we find
\be
\ba
&-(\Delta S_{A,0}^{(2)}+\Delta S_{A,\lambda}^{(2)})\\
=&\log\Big(\prod_{i=1}^{4}|\frac{dw_{i}}{dz_{i}}|^{-2h_{a}}\frac{\langle{\cal O}_{a}^{\dagger}(z_{1},\bar{z}_{1}){\cal O}_{a}(z_{2},\bar{z}_{2}){\cal O}_{a}^{\dagger}(z_{3},\bar{z}_{3}){\cal O}_{a}(z_{4},\bar{z}_{4})\rangle_{\Sigma_{1}}}{\langle{\cal O}_{a}^{\dagger}(w_{1},\bar{w}_{1}){\cal O}_{a}(w_{2},\bar{w}_{2})\rangle_{\Sigma_{1}}^{2}}\Big)\\
&-6\lambda\int d^{2}z\frac{c}{8\bar{z}^{2}}\frac{(L^{2}-\bar{z}^{2})^{2}}{2L^{3}\bar{z}}\frac{\langle J(z){\cal O}_{a}^{\dagger}(z_{1},\bar{z}_{1}){\cal O}_{a}(z_{2},\bar{z}_{2}){\cal O}_{a}^{\dagger}(z_{3},\bar{z}_{3}){\cal O}_{a}(z_{4},\bar{z}_{4})\rangle_{\Sigma_{1}}}{\langle{\cal O}_{a}^{\dagger}(z_{1},\bar{z}_{1}){\cal O}_{a}(z_{2},\bar{z}_{2}){\cal O}_{a}^{\dagger}(z_{3},\bar{z}_{3}){\cal O}_{a}(z_{4},\bar{z}_{4})\rangle_{\Sigma_{1}}}\\
&-6\lambda\int d^{2}z\frac{(L^{2}-\bar{z}^{2})^{2}}{2L^{3}\bar{z}}\frac{\langle J(z)\bar{T}(\bar{z}){\cal O}_{a}^{\dagger}(z_{1},\bar{z}_{1}){\cal O}_{a}(z_{2},\bar{z}_{2}){\cal O}_{a}^{\dagger}(z_{3},\bar{z}_{3}){\cal O}_{a}(z_{4},\bar{z}_{4})\rangle_{\Sigma_{1}}}{\langle{\cal O}_{a}^{\dagger}(z_{1},\bar{z}_{1}){\cal O}_{a}(z_{2},\bar{z}_{2}){\cal O}_{a}^{\dagger}(z_{3},\bar{z}_{3}){\cal O}_{a}(z_{4},\bar{z}_{4})\rangle_{\Sigma_{1}}}\\
&-6\lambda\int d^{2}w\frac{\langle J\bar{T}(w,\bar{w}){\cal O}_{a}^{\dagger}(w_{1},\bar{w}_{1}){\cal O}_{a}(w_{2},\bar{w}_{2})\rangle_{\Sigma_{1}}}{\langle{\cal O}_{a}^{\dagger}(w_{1},\bar{w}_{1}){\cal O}_{a}(w_{2},\bar{w}_{2})\rangle_{\Sigma_{1}}}.
\ea
\ee

We still focus on the large $c$ case, $\Delta S_{A,\lambda}^{(2)}$ can be written as
\be
\ba
\Delta S_{A,\lambda}^{(2)}&=6\lambda\int d^{2}z\frac{c}{8\bar{z}^{2}}\frac{(L^{2}-\bar{z}^{2})^{2}}{2L^{3}\bar{z}}\frac{\langle J(z){\cal O}_{a}^{\dagger}(z_{1},\bar{z}_{1}){\cal O}_{a}(z_{2},\bar{z}_{2}){\cal O}_{a}^{\dagger}(z_{3},\bar{z}_{3}){\cal O}_{a}(z_{4},\bar{z}_{4})\rangle_{\Sigma_{1}}}{\langle{\cal O}_{a}^{\dagger}(z_{1},\bar{z}_{1}){\cal O}_{a}(z_{2},\bar{z}_{2}){\cal O}_{a}^{\dagger}(z_{3},\bar{z}_{3}){\cal O}_{a}(z_{4},\bar{z}_{4})\rangle_{\Sigma_{1}}}+\cdots.\\
&=6\lambda\sum_{i=1}^{4}\int d^{2}z\frac{c}{8\bar{z}^{2}}\frac{(L^{2}-\bar{z}^{2})^{2}}{2L^{3}\bar{z}}\frac{q_{i}}{z-z_{i}}+{\cal O}(c^0),
\ea
\ee
where we used the Ward identity.
This integrals can be evaluated in the similar way as in Section \ref{sec:EE-TTbar}, and we find
we find
\be
\ba
\Delta S_{A,\lambda}^{(2)}&=6\lambda\frac{c}{8}\sum_{i=1}^{4}q_{i}\Big(-\frac{2\pi}{L}\log(\Lambda/|z_{i}|)+2\pi\big(-\frac{1}{4}\frac{|z_{i}|^{4}}{2L^{3}z_{i}^{2}}+\frac{1}{4}\frac{Lz_{i}^{2}}{2|z_{i}|^{4}}\big)\Big)+{\cal O}(c^0).
\ea
\ee
Plugging the coordinates (\ref{eq:z-coordinate}) in, we obtain
\be
\label{eq:REE-JTbar}
\Delta S_{A,\lambda}^{(2)}=6 \lambda\frac{c}{8} \frac{\pi\sum_{i=1}^{4}q_{i}}{4L}\Big(L\big(\frac{1}{l+L+t}+\frac{1}{l+t}\big)-2\log\big(\frac{(l+L-t)(l+L+t)}{l^{2}-t^{2}}\big)+8\log(\frac{L}{\Lambda})\Big)+{\cal O}(c^0,\lambda^2)
\ee
for $t>0$. From  the above equation, up to the $\lambda$ leading order of the $J\bar{T}$-deformation and the leading order of the large c limit, the R\'enyi entanglement entropy will obtain the corrections, where the first two terms associated with non trivial time dependence and the other term is about the UV cutoff due to the regularization.
We also find that the excess of R\'enyi entanglement entropy will be dramatically changed.

\subsection{OTOC in $J\bar{T}$-deformed CFTs}
We then consider the OTOC under the $J\bar{T}$-deformation. Under the coordinate transformation, the correlation function transform as
\be
\ba
&\langle J\bar{T}(w,\bar{w})W(w_{1},\bar{w}_{1})W(w_{2},\bar{w}_{2})V(w_{3},\bar{w}_{3})V(w_{4},\bar{w}_{4})\rangle_{\beta}\\
=&\prod_{i=1}^{4}(\frac{2\pi z_{i}}{\beta})^{h_{i}}(\frac{2\pi\bar{z}_{i}}{\beta})^{\bar{h}_{i}}(\frac{2\pi\bar{z}}{\beta})^{2}\frac{\partial z}{\partial w}\langle J(z)\big(\bar{T}(\bar{z})-\frac{c}{24\bar{z}^{2}}\big)W(z_{1},\bar{z}_{1})W(z_{2},\bar{z}_{2})V(z_{3},\bar{z}_{3})V(z_{4},\bar{z}_{4})\rangle.
\ea
\ee
The $J\bar{T}$-deformation of the function
\be
\frac{\langle W(w_{1},\bar{w}_{1})W(w_{2},\bar{w}_{2})V(w_{3},\bar{w}_{3})V(w_{4},\bar{w}_{4})\rangle_{\beta}}{\langle W(w_{1},\bar{w}_{1})W(w_{2},\bar{w}_{2})\rangle_{\beta}\langle V(w_{3},\bar{w}_{3})V(w_{4},\bar{w}_{4})\rangle_{\beta}}
\ee
becomes
\be
\ba
&\Big(\langle W(w_{1},\bar{w}_{1})W(w_{2},\bar{w}_{2})V(w_{3},\bar{w}_{3})V(w_{4},\bar{w}_{4})\rangle_{\beta}\\
&\qquad+\lambda\int d^{2}w_{a}\langle J\bar{T}(w_{a},\bar{w}_{a})W(w_{1},\bar{w}_{1})W(w_{2},\bar{w}_{2})V(w_{3},\bar{w}_{3})V(w_{4},\bar{w}_{4})\rangle_{\beta}\Big)\\
&\times\frac{1}{\Big(\langle W(w_{1},\bar{w}_{1})W(w_{2},\bar{w}_{2})\rangle_{\beta}+\lambda\int d^{2}w_{b}\langle J\bar{T}(w_{b},\bar{w}_{b})W(w_{1},\bar{w}_{1})W(w_{2},\bar{w}_{2})\rangle_{\beta}\Big)}\\
&\times\frac{1}{\Big(\langle V(w_{3},\bar{w}_{3})V(w_{4},\bar{w}_{4})\rangle_{\beta}+\lambda\int d^{2}w_{c}\langle J\bar{T}(w_{c},\bar{w}_{c})V(w_{3},\bar{w}_{3})V(w_{4},\bar{w}_{4})\rangle_{\beta}\Big)}\\
=&\frac{\langle W(w_{1},\bar{w}_{1})W(w_{2},\bar{w}_{2})V(w_{3},\bar{w}_{3})V(w_{4},\bar{w}_{4})\rangle_{\beta}}{\langle W(w_{1},\bar{w}_{1})W(w_{2},\bar{w}_{2})\rangle_{\beta}\langle V(w_{3},\bar{w}_{3})V(w_{4},\bar{w}_{4})\rangle_{\beta}}\\
&\times\Big(1-\lambda\int d^{2}w_{b}\frac{\langle J\bar{T}(w_{b},\bar{w}_{b})W(w_{1},\bar{w}_{1})W(w_{2},\bar{w}_{2})\rangle_{\beta}}{\langle W(w_{1},\bar{w}_{1})W(w_{2},\bar{w}_{2})\rangle_{\beta}}-\lambda\int d^{2}w_{c}\frac{\langle J\bar{T}(w_{c},\bar{w}_{c})V(w_{3},\bar{w}_{3})V(w_{4},\bar{w}_{4})\rangle_{\beta}}{\langle V(w_{3},\bar{w}_{3})V(w_{4},\bar{w}_{4})\rangle_{\beta}}\\
&+\lambda\frac{\int d^{2}w_{a}\langle J\bar{T}(w_{a},\bar{w}_{a})W(w_{1},\bar{w}_{1})W(w_{2},\bar{w}_{2})V(w_{3},\bar{w}_{3})V(w_{4},\bar{w}_{4})\rangle_{\beta}}{\langle W(w_{1},\bar{w}_{1})W(w_{2},\bar{w}_{2})V(w_{3},\bar{w}_{3})V(w_{4},\bar{w}_{4})\rangle_{\beta}}\\&+{\cal O}(\lambda^{2})\Big),
\ea
\ee
where we have expanded around $\lambda=0$. Under the conformal transformation (\ref{eq:conf-map-OTOC}), we obtain
\be
\ba
&\frac{\langle W(w_{1},\bar{w}_{1})W(w_{2},\bar{w}_{2})V(w_{3},\bar{w}_{3})V(w_{4},\bar{w}_{4})\rangle_{\beta}}{\langle W(w_{1},\bar{w}_{1})W(w_{2},\bar{w}_{2})\rangle_{\beta}\langle V(w_{3},\bar{w}_{3})V(w_{4},\bar{w}_{4})\rangle_{\beta}}\\
&\times\Big(1+\lambda\frac{c}{24}\frac{2\pi}{\beta}\Big[-\int d^{2}z\frac{1}{\bar{z}}\frac{\langle J(z)W(z_{1},\bar{z}_{1})W(z_{2},\bar{z}_{2})V(z_{3},\bar{z}_{3})V(z_{4},\bar{z}_{4})\rangle}{\langle W(z_{1},\bar{z}_{1})W(z_{2},\bar{z}_{2})V(z_{3},\bar{z}_{3})V(z_{4},\bar{z}_{4})}\\
&+\lambda\frac{c}{24}\int d^{2}z\frac{1}{\bar{z}}\frac{\langle J(z)W(z_{1},\bar{z}_{1})W(z_{2},\bar{z}_{2})\rangle}{\langle W(z_{1},\bar{z}_{1})W(z_{2},\bar{z}_{2})\rangle}+\lambda\frac{c}{24}\int d^{2}z\frac{1}{\bar{z}}\frac{\langle J(z)V(z_{3},\bar{z}_{3})V(z_{4},\bar{z}_{4})\rangle}{\langle V(z_{3},\bar{z}_{3})V(z_{4},\bar{z}_{4})}\Big]\Big)\\
&+\lambda\frac{\int d^{2}z\frac{2\pi\bar{z}}{\beta}\langle J(z)\bar{T}(\bar{z})W(z_{1},\bar{z}_{1})W(z_{2},\bar{z}_{2})V(z_{3},\bar{z}_{3})V(z_{4},\bar{z}_{4})\rangle}{\langle W(z_{1},\bar{z}_{1})W(z_{2},\bar{z}_{2})V(z_{3},\bar{z}_{3})V(z_{4},\bar{z}_{4})}\\
&-\lambda\frac{\int d^{2}z\frac{2\pi\bar{z}}{\beta}\langle J(z)\bar{T}(\bar{z})W(z_{1},\bar{z}_{1})W(z_{2},\bar{z}_{2})\rangle}{\langle W(z_{1},\bar{z}_{1})W(z_{2},\bar{z}_{2})\rangle}-\lambda\frac{\int d^{2}z\frac{2\pi\bar{z}}{\beta}\langle J(z)\bar{T}(\bar{z})V(z_{3},\bar{z}_{3})V(z_{4},\bar{z}_{4})\rangle}{\langle V(z_{3},\bar{z}_{3})V(z_{4},\bar{z}_{4})}.
\ea
\ee
By using the Ward identity, we find
\be
\ba
&\frac{\langle W(w_{1},\bar{w}_{1})W(w_{2},\bar{w}_{2})V(w_{3},\bar{w}_{3})V(w_{4},\bar{w}_{4})\rangle_{\beta}}{\langle W(w_{1},\bar{w}_{1})W(w_{2},\bar{w}_{2})\rangle_{\beta}\langle V(w_{3},\bar{w}_{3})V(w_{4},\bar{w}_{4})\rangle_{\beta}}\\
&\Big(1+\lambda\frac{c}{24}\frac{2\pi}{\beta}\Big[-\int d^{2}z\frac{1}{\bar{z}}\sum_{i=1}^{4}\frac{q_{i}}{z-z_{i}}
+\lambda\frac{c}{24}\int d^{2}z\frac{1}{\bar{z}}\sum_{i=1}^{2}\frac{q_{i}}{z-z_{i}}+\lambda\frac{c}{24}\int d^{2}z\frac{1}{\bar{z}}\sum_{i=3}^{4}\frac{q_{i}}{z-z_{i}}\Big]\\
&+\lambda\int d^{2}z\frac{2\pi\bar{z}}{\beta}\Big(\sum_{i=1}^{4}\frac{q_{i}}{z-z_{i}}\Big)\Big(h_{w}\frac{\bar{z}_{12}^{2}}{(\bar{z}-\bar{z}_{1})^{2}(\bar{z}-\bar{z}_{2})^{2}}+h_{v}\frac{\bar{z}_{34}^{2}}{(\bar{z}-\bar{z}_{3})^{2}(\bar{z}-\bar{z}_{4})^{2}}+\frac{\bar{z}_{14}\bar{z}_{23}}{\prod_{i=1}^{4}(\bar{z}-\bar{z}_{i})}\frac{\bar{\eta}\partial_{\bar{\eta}}G(\eta,\eta)}{G(\eta,\eta)}\Big)\\&-\lambda\int d^{2}z\frac{2\pi\bar{z}}{\beta}\Big(\frac{q_{1}}{z-z_{1}}+\frac{q_{2}}{z-z_{2}}\Big)\frac{h_{w}(\bar{z}_{1}-\bar{z}_{2})^{2}}{(\bar{z}-\bar{z}_{1})^{2}(\bar{z}-\bar{z}_{2})^{2}}\\&-\lambda\int d^{2}z\frac{2\pi\bar{z}}{\beta}\Big(\frac{q_{3}}{z-z_{3}}+\frac{q_{4}}{z-z_{4}}\Big)\frac{h_{v}(\bar{z}_{3}-\bar{z}_{4})^{2}}{(\bar{z}-\bar{z}_{3})^{2}(\bar{z}-\bar{z}_{4})^{2}}
+{\cal O}(\lambda^2)\Big)\\
=&\frac{\langle W(w_{1},\bar{w}_{1})W(w_{2},\bar{w}_{2})V(w_{3},\bar{w}_{3})V(w_{4},\bar{w}_{4})\rangle_{\beta}}{\langle W(w_{1},\bar{w}_{1})W(w_{2},\bar{w}_{2})\rangle_{\beta}\langle V(w_{3},\bar{w}_{3})V(w_{4},\bar{w}_{4})\rangle_{\beta}}\Big(1\\
&+\lambda\int d^{2}z\frac{2\pi\bar{z}}{\beta}\Big\{\Big(\frac{q_{3}}{z-z_{3}}+\frac{q_{4}}{z-z_{4}}\Big)h_{w}\frac{\bar{z}_{12}^{2}}{(\bar{z}-\bar{z}_{1})^{2}(\bar{z}-\bar{z}_{2})^{2}}+\Big(\frac{q_{1}}{z-z_{1}}+\frac{q_{2}}{z-z_{2}}\Big)h_{v}\frac{\bar{z}_{34}^{2}}{(\bar{z}-\bar{z}_{3})^{2}(\bar{z}-\bar{z}_{4})^{2}}\\
&+\Big(\sum_{i=1}^{4}\frac{q_{i}}{z-z_{i}}\Big)\frac{\bar{z}_{14}\bar{z}_{23}}{\prod_{i=1}^{4}(\bar{z}-\bar{z}_{i})}\frac{\bar{\eta}\partial_{\bar{\eta}}G(\eta,\eta)}{G(\eta,\eta)}\Big\}+{\cal O}(\lambda^2)\Big).
\ea
\ee
This integral can be evaluated by using the similar method as in the Section \ref{sec:EE-TTbar}. Using the coordinates (\ref{eq:OTOC-z}) and conformal block (\ref{eq:CB-F}), we obtain
\be
\ba
&\quad \lambda\int d^{2}z \bar{z}\Big\{\Big(\frac{q_{3}}{z-z_{3}}+\frac{q_{4}}{z-z_{4}}\Big)h_{w}\frac{\bar{z}_{12}^{2}}{(\bar{z}-\bar{z}_{1})^{2}(\bar{z}-\bar{z}_{2})^{2}}+\Big(\frac{q_{1}}{z-z_{1}}+\frac{q_{2}}{z-z_{2}}\Big)h_{v}\frac{\bar{z}_{34}^{2}}{(\bar{z}-\bar{z}_{3})^{2}(\bar{z}-\bar{z}_{4})^{2}}\\
&=\lambda\frac{\pi}{2}(q_{3}+q_{4})e^{\frac{4\pi x}{\beta}}\frac{\big(e^{\frac{8\pi(t+i\epsilon_{3})}{\beta}}-1\big)}{e^{\frac{4\pi(t+i\epsilon_{3})}{\beta}}+e^{\frac{4\pi(t+2x+i\epsilon_{3})}{\beta}}-e^{\frac{4\pi(2t+x+2i\epsilon_{3})}{\beta}}-e^{\frac{4\pi x}{\beta}}},
\ea
\ee
where we have located the operators in pairs: $\epsilon_2=\epsilon_1+\beta/2$ and $\epsilon_4=\epsilon_3+\beta/2$, and set $\epsilon_1=0$ without loss generality \cite{Perlmutter:2016pkf}. { At late time, the $J\bar{T}$-defomed  OTOC behaves as
\be
\ba
&\frac{\langle W(w_{1},\bar{w}_{1})W(w_{2},\bar{w}_{2})V(w_{3},\bar{w}_{3})V(w_{4},\bar{w}_{4})\rangle_{\beta}}{\langle W(w_{1},\bar{w}_{1})W(w_{2},\bar{w}_{2})\rangle_{\beta}\langle V(w_{3},\bar{w}_{3})V(w_{4},\bar{w}_{4})\rangle_{\beta}}\\
\xrightarrow{J\bar{T}}  &\frac{\langle W(w_{1},\bar{w}_{1})W(w_{2},\bar{w}_{2})V(w_{3},\bar{w}_{3})V(w_{4},\bar{w}_{4})\rangle_{\beta}}{\langle W(w_{1},\bar{w}_{1})W(w_{2},\bar{w}_{2})\rangle_{\beta}\langle V(w_{3},\bar{w}_{3})V(w_{4},\bar{w}_{4})\rangle_{\beta}}\Big(1+\lambda C_3(x)e^{-\frac{2\pi}{\beta}t}+\cdots\Big),
\ea
\ee
where $C_3(x)$ is a coefficient independent of $t$. We thus find the $J\bar{T}$-deformation does not affect the maximal chaos found by OTOC up to the perturbation first order of the deformation}, which is the similar as the result found in $T\bar{T}$-deformation.

\section{Conclusions and discussions}
In this paper, we study the $T\bar{T}$/$J\bar T$-deformation of two-dimensional CFTs perturbatively in the first order of the deformation. Thanks to the energy momentum conservation and current conservation, we employ the Ward identity to study the 2- and 4-point correlation functions in a perturbative manner. To obtain the closed form for these correlation functions, we make use of dimensional regularization to deal with the space time integral. Our results can exactly reproduce the previous results related to 2-point correlation of deformed CFTs in the literature \cite{Kraus:2018xrn, Guica:2019vnb, Cardy:2019qao}. As an application of this work, we study the R\'enyi entanglement entropy of the locally excited state in deformed CFTs. With such deformation, at the leading order ${\cal O}(c^2)$, the R\'enyi entanglement entropy depends on the UV and IR cut off introduced by the regularization.
At the order ${\cal O}(c)$ the R\'enyi entanglement entropy acquires a non-trivial time dependence. The excess of R\'enyi entanglement entropy between early and late times is significantly changed up to the order ${\cal O}(c)$, see (\ref{eq:Ex-REE}) and (\ref{eq:REE-JTbar}).

In \cite{Smirnov:2016lqw}, it claimed that the integrability structure is still held in integrable models with $T\bar{T}$-deformation. We read out the signals of integrability by calculating the OTOC in the $T\bar{T}$/$J\bar T$-deformed field theory. To this end, the OTOC of deformed theory has been given explicitly and it shows that the $T\bar{T}$/$J\bar T$-deformation does not change the maximal chaotic property of holographic CFTs in our calculation. Although we do not explicitly exhibit the integrability structure of $T\bar{T}$/$J\bar T$-deformed integrable CFTs, up to the first order of deformation, we expect that such deformations do not change the integrability structure of un-deformed theory which is an interesting direction in the future work.

One can directly extend the perturbation to the higher order of these deformations to calculate the higher-point correlation functions, which will give us some highly non-trivial insights into the renormalization flow structure of the correlation function. One can compare the correlation function in the deformed theory with the non-perturbative correlation functions proposed by \cite{Cardy:2019qao}, and check the non-local effect in the UV limit. Further, one can exactly check the crossing symmetry of four-point function in a perturbative sense, as we have done in this paper,  or non-perturbative sense \cite{Cardy:2019qao} as elsewhere. To exactly match these two methods is a very interesting direction for future research. As applications, one can apply these higher order corrected correlation functions to study the R\'enyi entanglement entropy and the OTOC to see the chaotic signals of the deformed theory in a perturbative sense.

\section*{Acknowledgements}
The authors are grateful to Bin Chen, Axel Kleinschmidt, Tadashi Takayanagi, Stefan Theisen, Yuan Sun and Jiaju Zhang for useful discussions.
SH thanks the Yukawa Institute for Theoretical Physics at Kyoto University.
Discussions during the workshop YITP-T-19-03 "Quantum Information and String Theory 2019"
were useful to complete this work. SH would like to thank Department of Physics in Tokyo Institute of Technology for warm hospitality during the process of this project.  SH also would like to appreciate the financial support from Jilin University and Max Planck Partner group. HS would like to thank Jilin University, South China Normal University, Sun Yat-Sen University and Korea Institute for Advanced Study for warm hospitality.

\appendix

\section{Useful integrals}
\subsection{Notation of the integrals}
It is convenient to use the notation:
\be
\label{eq:integrals}
{\cal I}_{a_1,\cdots, a_m, b_1,\cdots, b_n}(z_{i_1},\cdots,z_{i_m}, \bar{z}_{j_1},\cdots, \bar{z}_{j_n}):=\int \frac{ d^2z}{(z-z_{i_1})^{a_1}\cdots (z-z_{i_m})^{a_m} (\bar{z}-\bar{z}_{j_1})^{b_1}\cdots (\bar{z}-\bar{z}_{j_n})^{b_n}}.
\ee
For examples, we write
\be
\ba
{\cal I}_{2222}(z_{1},z_{2}, \bar{z}_{1}, \bar{z}_{2})&=\int\frac{d^{2}z}{|z-z_{1}|^{4}|z-z_{2}|^{4}}\\
{\cal I}_{2222}(z_{1},z_{2},\bar{z}_{3},\bar{z}_{4})&=\int\frac{d^{2}z}{(z-z_{1})^{2}(z-z_{2})^{2}(\bar{z}-\bar{z}_{3})^{2}(\bar{z}-\bar{z}_{4})^{2}}\\
{\cal I}_{221111}(z_{1},z_{2},\bar{z}_{1},\bar{z}_{2},\bar{z}_{3},\bar{z}_{4})&=\int\frac{d^{2}z}{(z-z_{1})^{2}(z-z_{2})^{2}(\bar{z}-\bar{z}_{1})(\bar{z}-\bar{z}_{2})(\bar{z}-\bar{z}_{3})(\bar{z}-\bar{z}_{4})}\\
{\cal I}_{11111111}(z_{1},z_{2},z_{3},z_{4},\bar{z}_{1},\bar{z}_{2},\bar{z}_{3},\bar{z}_{4})&=\int \frac{d^{2}z}{\prod_{i=1}^{4}(z-z_{i})\prod_{j=1}^{4}(\bar{z}-\bar{z}_{j})}.
\ea
\ee
Moreover, we will also write
\be
\ba
{\cal I}_1(z_1,z_2)&={\cal I}_{1111}(z_1,z_2,\bar{z}_1,\bar{z}_2)\\
{\cal I}_2(z_1,z_2)&={\cal I}_{2222}(z_1,z_2,\bar{z}_1,\bar{z}_2)\\
{\cal I}_3(z_1,\bar{z}_2)&={\cal I}_{11}(z_1,\bar{z}_2),
\ea
\ee
which is used to expressed other more complicated integrals. The formula of ${\cal I}_{2}(z_{1},z_{2})$ and ${\cal I}_{2222}(z_{1},z_{2},\bar{z}_{3},\bar{z}_{4})$ can be found in (\ref{eq:I2-regulated}) and (\ref{eq:I2222z1-z2bz3bz4}) respectively. The formula of ${\cal I}_{221111}(z_{1},z_{2},\bar{z}_{1},\bar{z}_{2},\bar{z}_{3},\bar{z}_{4})$ and ${\cal I}_{11111111}(z_{1},z_{2},z_{3},z_{4},\bar{z}_{1},\bar{z}_{2},\bar{z}_{3},\bar{z}_{4})$ can be found in (\ref{eq:I221111}) and (\ref{eq:I11111111}) respectively.

\subsubsection{${\cal I}_1(z_{1},z_{2})={\cal I}_{1111}(z_{1},z_{2}, \bar{z}_{1}, \bar{z}_{2})$}
Let us first consider the integrals
\be
{\cal I}_1(z_{1},z_{2})={\cal I}_{1111}(z_{1},z_{2}, \bar{z}_{1}, \bar{z}_{2})=\int\frac{d^{2}z}{|z-z_{1}|^{2}|z-z_{2}|^{2}},
\ee
which can be performed by introducing a Feynman parameter:
\be
\ba
{\cal I}_{1}(z_{1},z_{2})=\int_{0}^{1}du\int\frac{d^{2}\tilde{z}}{\Big(|\tilde{z}|^{2}+u(1-u)|z_{12}|^{2}\Big)^{2}}=2V_{S^{2-1}}\int_{0}^{1}du\int_{0}^{\infty}\frac{\rho^{2-1}d\rho}{\Big(\rho^{2}+u(1-u)|z_{12}|^{2}\Big)^{2}}\ea
\ee
where $z=\tilde{z}+uz_{1}+(1-u)z_{2}$, $\bar{z}=\bar{\tilde{z}}+u\bar{z}_{1}+(1-u)\bar{z}_{2}$. To regulate the divergence, we use the dimensional regularization by replacing two-dimensional to $d$-dimensiuon:
\be
\ba
{\cal I}_{1}^{(d)}(z_{1},z_{2})&=2V_{S^{d-1}}\int_{0}^{1}du\int_{0}^{\infty}\frac{\rho^{d-1}d\rho}{\Big(\rho^{2}+u(1-u)|z_{12}|\Big)^{2}}\\
&\xrightarrow{d=2+\epsilon}\frac{2\pi}{|z_{12}|^{2}}\Big(\frac{2}{\epsilon}+\log|z_{12}|^{2}+\gamma+\log\pi+{\cal O}(\epsilon)\Big)\label{eq:I1-dim-reg}
\ea
\ee
where $\epsilon>0$. We have used $V_{S^{d-1}}=\frac{{\pi^{\frac{d}{2}}}}{\Gamma(\frac{d}{2})}$.

\subsubsection{${\cal I}_2(z_{1},z_{2})={\cal I}_{2222}(z_{1},z_{2}, \bar{z}_{1}, \bar{z}_{2})$}
By using the Feynman parameter
\be
\frac{1}{A_{1}^{2}A_{2}^{2}}=\frac{\Gamma(4)}{\Gamma(2)\Gamma(2)}\int_{0}^{1}du\frac{u(1-u)}{\Big(uA_{1}+(1-u)A_{2}\Big)^{4}},
\ee
the second integrals can be written by
\be
\ba
{\cal I}_{2}(z_{1},z_{2})&={\cal I}_{2222}(z_{1},z_{2}, \bar{z}_{1}, \bar{z}_{2})=\int dz^{2}\frac{1}{\Big(|z-z_{1}|^{2}|z-z_{2}|^{2}\Big)^{2}}
\\&
=6\int_{0}^{1}du\int d\tilde{z}^{2}\frac{u(1-u)}{\Big(|\tilde{z}|^{2}+u(1-u)|z_{12}|^{2}\Big)^{4}}\\&=12V_{S^{2-1}}\int_{0}^{1}duu(1-u)\int_{0}^{\infty}\frac{\rho^{2-1}d\rho}{\Big(\rho^{2}+u(1-u)|z_{12}|^{2}\Big)^{4}}.
\ea
\ee
To regulate the divergence, we use the dimensional regularization by 
\be
\ba
\label{eq:I2-regulated}
{\cal I}_{2}^{(d)}(z_{1},z_{2})&=12V_{S^{d-1}}\int_{0}^{1}duu(1-u)\int_{0}^{\infty}\frac{\rho^{d-1}d\rho}{\Big(\rho^{2}+u(1-u)|z_{12}|^{2}\Big)^{4}}\\
&\xrightarrow{d=2+\epsilon}\frac{{4}\pi}{|z_{12}|^{6}}\Big(\frac{4}{\epsilon}+2\log|z_{12}|^{2}+2\log\pi+2\gamma-5\Big).
\ea
\ee

\subsubsection{${\cal I}_3(z_{1},z_{2})={\cal I}_{11}(z_1,\bar{z}_2)$}
We then consider
\be
{\cal I}_{3}(z_{1}, \bar{z}_2)={\cal I}_{11}(z_1,\bar{z}_2)=\int d^{2}z\frac{(\bar{z}-\bar{z}_{1})(z-z_{2})}{|z-z_{1}|^{2}|z-z_{2}|^{2}}.
\ee
By using the Feynman parameter
\be
\ba
{\cal I}_{3}(z_{1}, \bar{z}_{2})
&=\int_{0}^{1}du\int d^{2}z\frac{(\bar{z}-\bar{z}_{1})(z-z_{2})}{\Big(u|z-z_{1}|^{2}+(1-u)|z-z_{2}|^{2}\Big)^{2}}\\&=\int_{0}^{1}du\int d^{2}z\frac{(\tilde{\bar{z}}-(1-u)\bar{z}_{12})(\tilde{z}+uz_{12})}{\Big(|\tilde{z}|^{2}+u(1-u)|z_{12}|^{2}\Big)^{2}}\\
&=\int_{0}^{1}du\int d^{2}z\frac{\rho^{2}-u(1-u)|z_{12}|^{2}}{\Big(\rho^{2}+u(1-u)|z_{12}|^{2}\Big)^{2}}.%
\ea
\ee

We then replace two-dimension to $d$-dimension:
\be
\ba
\label{eq:I3-regulated}
{\cal I}_{3}^{(d)}(z_{1}, \bar{z}_{2})&=2V_{S^{d-1}}\int_{0}^{1}du\int_{0}^{\infty}d\rho\rho^{d-1}\frac{\rho^{2}-u(1-u)|z_{12}|^{2}}{\Big(\rho^{2}+u(1-u)|z_{12}|^{2}\Big)^{2}}\\&
\xrightarrow{d=2+\tilde{\epsilon}}-\pi\Big(\frac{2}{\tilde{\epsilon}}+\log|z_{12}|^{2}+\log\pi+\gamma\Big)+{\cal O}(\tilde{\epsilon}),
\ea
\ee
where $\tilde{\epsilon}<0$.

\subsection{Useful integrals for four-point function}
\subsubsection{${\cal I}_{2222}(z_{1},z_{2},\bar{z}_{3},\bar{z}_{4})$}
One may use the integrals (\ref{eq:I3-regulated}) to compute other more complicated integrals.
It is easy to see
\be
\ba
\label{eq:I4-122}
{\cal I}_{122}(z_{1},\bar{z}_{3},\bar{z}_{4})&=\int d^{2}z\frac{1}{(z-z_{1})(\bar{z}-\bar{z}_{3})^{2}(\bar{z}-\bar{z}_{4})^{2}}\\
&=\partial_{\bar{z}_{4}}\partial_{\bar{z}_{3}}\Big(\frac{1}{\bar{z}_{34}}\int d^{2}z\big(\frac{1}{(z-z_{1})(\bar{z}-\bar{z}_{3})}-\frac{1}{(z-z_{1})(\bar{z}-\bar{z}_{4})}\big)\Big)\\
&=\partial_{\bar{z}_{4}}\partial_{\bar{z}_{3}}\Big(\frac{1}{\bar{z}_{34}}\big({\cal I}_{3}(z_{1},\bar{z}_{3})-{\cal I}_{3}(z_{1},\bar{z}_{4})\big)\Big).
\ea
\ee
Other useful integral is
\be
\ba
\label{eq:I122z1bz1bz3}
\quad {\cal I}_{122}(z_{1},\bar{z}_{1},\bar{z}_{3})&=\int d^{2}z\frac{1}{(z-z_{1})(\bar{z}-\bar{z}_{1})^{2}(\bar{z}-\bar{z}_{3})^{2}}\\
&=\int d^{2}z\frac{1}{(z-z_{1})}\frac{1}{\bar{z}_{13}^{2}}\Big(\frac{1}{(\bar{z}-\bar{z}_{1})^{2}}+\frac{1}{(\bar{z}-\bar{z}_{3})^{2}}-\frac{2}{(\bar{z}-\bar{z}_{1})(\bar{z}-\bar{z}_{3})}\Big)\\
&=\frac{1}{\bar{z}_{13}^{2}}\Big(-\frac{2}{\bar{z}_{13}}{\cal I}_{11}(z_{1},\bar{z}_{1})+\partial_{\bar{z}_{3}}{\cal I}_{3}(z_{1},\bar{z}_{3})+\frac{2}{\bar{z}_{13}}{\cal I}_{3}(z_{1},\bar{z}_{3})\Big),
\ea
\ee
where we used ${\cal I}_{21}(z_1,\bar{z_1})=0$. From (\ref{eq:I4-122}) and (\ref{eq:I122z1bz1bz3}), we can write
\be
\ba
\label{eq:I2222z1-z2bz3bz4}
{\cal I}_{2222}(z_{1},z_{2},\bar{z}_{3},\bar{z}_{4})&:=\int\frac{d^{2}z}{(z-z_{1})^{2}(z-z_{2})^{2}(\bar{z}-\bar{z}_{3})^{2}(\bar{z}-\bar{z}_{4})^{2}}\\
&=\partial_{z_{2}}\partial_{z_{1}}\Big(\frac{1}{z_{12}}\int\frac{d^{2}z}{(\bar{z}-\bar{z}_{3})^{2}(\bar{z}-\bar{z}_{4})^{2}}\big(\frac{1}{z-z_{1}}-\frac{1}{z-z_{2}}\big)\Big)\\
&=\partial_{z_{2}}\partial_{z_{1}}\Big(\frac{1}{z_{12}}\big({\cal I}_{122}(z_{1},\bar{z}_{3},\bar{z}_{4})-{\cal I}_{122}(z_{2},\bar{z}_{3},\bar{z}_{4})\big)\Big).
\ea
\ee
Therefore, we can express ${\cal I}_{2222}(z_{1},z_{2},\bar{z}_{3},\bar{z}_{4})$ by using ${\cal I}_3$.

\subsubsection{${\cal I}_{221111}(z_{1},z_{2},\bar{z}_{1},\bar{z}_{2},\bar{z}_{3},\bar{z}_{4})$}
Let us consider how to compute the integrals
\be
{\cal I}_{221111}(z_{1},z_{2},\bar{z}_{1},\bar{z}_{2},\bar{z}_{3},\bar{z}_{4})=\int\frac{d^{2}z}{(z-z_{1})^{2}(z-z_{2})^{2}(\bar{z}-\bar{z}_{1})(\bar{z}-\bar{z}_{2})(\bar{z}-\bar{z}_{3})(\bar{z}-\bar{z}_{4})}.
\ee
As preparation, we first consider the integrals ${\cal I}_{221}(z_{1},z_{2},\bar{z}_{3})$ and ${\cal I}_{221}(z_{1},z_{2},\bar{z}_{1})$. By using $\partial_{z_{1}}\partial_{z_{2}}\Big(\frac{1}{z_{12}}\big(\frac{1}{z-z_{1}}-\frac{1}{z-z_{2}}\big)\Big)=\frac{1}{(z-z_{1})^{2}(z-z_{2})^{2}}$, it is easy to find
\be
\ba
{\cal I}_{221}(z_{1},z_{2},\bar{z}_{3})
&=\partial_{z_{1}}\partial_{z_{2}}\Big(\frac{1}{z_{12}}\int d^{2}z\frac{1}{(\bar{z}-\bar{z}_{3})}\big(\frac{1}{z-z_{1}}-\frac{1}{z-z_{2}}\big)\Big)\\&=\partial_{z_{1}}\partial_{z_{2}}\Big(\frac{1}{z_{12}}\big({\cal I}_{3}(z_{1},\bar{z}_{3})-{\cal I}_{3}(z_{2},\bar{z}_{3})\big)\Big).
\ea
\ee
Using $\frac{1}{(z-z_i)(z-z_j)}=\frac{1}{z_{ij}}\big(\frac{1}{z-z_i}-\frac{1}{z-z_j}\big)$ repeatly, we find
\be
\ba
{\cal I}_{221}(z_{1},z_{2},\bar{z}_{1})
&=\int\frac{d^{2}z}{(z-z_{1})^{2}(z-z_{2})^{2}(\bar{z}-\bar{z}_{1})}\\
&=\frac{1}{z_{12}^{2}}\int\frac{d^{2}z}{(\bar{z}-\bar{z}_{1})}\Big(\frac{1}{(z-z_{1})^{2}}+\frac{1}{(z-z_{2})^{2}}-\frac{2}{(z-z_{1})(z-z_{2})}\Big)\\
&=\frac{1}{z_{12}^{2}}\partial_{z_{2}}{\cal I}_{3}(z_{2},\bar{z}_{1})-\frac{2}{z_{12}^{3}}({\cal I}_{11}(z_{1},\bar{z}_{1})-{\cal I}_{3}(z_{2},\bar{z}_{1})),
\ea
\ee
where we used the factor that ${\cal I}_{11}(z_1,\bar{z_1})=0$. Since
\be
\ba
&\frac{1}{(z-z_{1})^{2}(z-z_{2})^{2}(\bar{z}-\bar{z}_{1})(\bar{z}-\bar{z}_{2})(\bar{z}-\bar{z}_{3})(\bar{z}-\bar{z}_{4})}\\
=&\frac{1}{\bar{z}_{12}\bar{z}_{34}}\frac{1}{(z-z_{1})^{2}(z-z_{2})^{2}}\Big(\frac{1}{(\bar{z}-\bar{z}_{1})}-\frac{1}{(\bar{z}-\bar{z}_{2})}\Big)\Big(\frac{1}{(\bar{z}-\bar{z}_{3})}-\frac{1}{(\bar{z}-\bar{z}_{4})}\Big).
\ea
\ee
We thus can express ${\cal I}_{221111}(z_{1},z_{2},\bar{z}_{1},\bar{z}_{2},\bar{z}_{3},\bar{z}_{4})$ as
\be
\ba
\label{eq:I221111}
&\quad{\cal I}_{221111}(z_{1},z_{2},\bar{z}_{1},\bar{z}_{2},\bar{z}_{3},\bar{z}_{4})\\
=&\frac{1}{\bar{z}_{12}\bar{z}_{34}}\Big({\cal I}_{2211}(z_{1},z_{2},\bar{z}_{1},\bar{z}_{3})-{\cal I}_{2211}(z_{2},z_{1},\bar{z}_{2},\bar{z}_{3})-{\cal I}_{2211}(z_{1},z_{2},\bar{z}_{1},\bar{z}_{4})+{\cal I}_{2211}(z_{2},z_{1},\bar{z}_{2},\bar{z}_{4})\Big)\\
=&\frac{1}{\bar{z}_{12}\bar{z}_{34}}\Big(\frac{1}{\bar{z}_{13}}\big({\cal I}_{221}(z_{1},z_{2},\bar{z}_{1})-{\cal I}_{221}(z_{1},z_{2},\bar{z}_{3})\big)-\frac{1}{\bar{z}_{23}}\big({\cal I}_{221}(z_{2},z_{1},\bar{z}_{2})-{\cal I}_{221}(z_{2},z_{1},\bar{z}_{3})\big)\\
&-\frac{1}{\bar{z}_{14}}\big({\cal I}_{221}(z_{1},z_{2},\bar{z}_{1})-{\cal I}_{221}(z_{1},z_{2},\bar{z}_{4})\big)+\frac{1}{\bar{z}_{24}}\big({\cal I}_{221}(z_{2},z_{1},\bar{z}_{2})-{\cal I}_{221}(z_{2},z_{1},\bar{z}_{4})\big)\\
\ea
\ee

The other important formula is
\be
\ba
\label{eq:I111122}
&\quad {\cal I}_{111122}(z_{1},z_{2},z_{3},z_{4},\bar{z}_{1},\bar{z}_{2})\\
&=\int\frac{d^{2}z}{(z-z_{1})(z-z_{2})(z-z_{3})(z-z_{4})(\bar{z}-\bar{z}_{1})^{2}(\bar{z}-\bar{z}_{2})^{2}}\\
&=\frac{1}{z_{12}z_{34}}\Big(\frac{{\cal I}_{122}(z_{1},\bar{z}_{1},\bar{z}_{2})-{\cal I}_{122}(z_{3},\bar{z}_{1},\bar{z}_{2})}{z_{13}}-\frac{{\cal I}_{122}(z_{2},\bar{z}_{1},\bar{z}_{2})-{\cal I}_{122}(z_{3},\bar{z}_{1},\bar{z}_{2})}{z_{23}}\\
&\quad-\frac{{\cal I}_{122}(z_{1},\bar{z}_{1},\bar{z}_{2})-{\cal I}_{122}(z_{4},\bar{z}_{1},\bar{z}_{2})}{z_{14}}+\frac{{\cal I}_{122}(z_{2},\bar{z}_{1},\bar{z}_{2})-{\cal I}_{122}(z_{4},\bar{z}_{1},\bar{z}_{2})}{z_{24}}\Big).
\ea
\ee

\subsubsection{${\cal I}_{11111111}(z_{1},z_{2},z_{3},z_{4},\bar{z}_{1},\bar{z}_{2},\bar{z}_{3},\bar{z}_{4})$}
It is very useful to factorize the complicated integral in terms of simple ones by using
\be
\frac{1}{(z-z_i)(z-z_j)}=\frac{1}{z_{ij}}\big(\frac{1}{z-z_i}-\frac{1}{z-z_j}\big),\quad \frac{1}{(\bar{z}-\bar{z}_i)(\bar{z}-\bar{z}_j)}=\frac{1}{\bar{z}_{ij}}\big(\frac{1}{\bar{z}-\bar{z}_i}-\frac{1}{\bar{z}-\bar{z}_j}\big).
\ee
For examples, ${\cal I}_{1111}(z_{1},z_{2},\bar{z}_{3},\bar{z}_{4})$ and ${\cal I}_{1111}(z_{1},z_{3},\bar{z}_{1},\bar{z}_{2})$ are evaluated as
\be
\ba
\label{eq:I1111z1234}
&\quad {\cal I}_{1111}(z_{1},z_{2},\bar{z}_{3},\bar{z}_{4})\\
&=\int d^{2}z\frac{1}{(z-z_{1})(z-z_{2})(\bar{z}-\bar{z}_{3})(\bar{z}-\bar{z}_{4})}\\
&=\frac{1}{z_{12}\bar{z}_{34}}\int d^{2}z\big(\frac{1}{(z-z_{1})}-\frac{1}{(z-z_{2})}\big)\big(\frac{1}{(\bar{z}-\bar{z}_{3})}-\frac{1}{(\bar{z}-\bar{z}_{4})}\big)\\
&=\frac{1}{z_{12}\bar{z}_{34}}\int d^{2}z\big(\frac{1}{(z-z_{1})(\bar{z}-\bar{z}_{3})}-\frac{1}{(z-z_{2})(\bar{z}-\bar{z}_{3})}-\frac{1}{(z-z_{1})(\bar{z}-\bar{z}_{4})}+\frac{1}{(z-z_{2})(\bar{z}-\bar{z}_{4})}\big)\\
&=\frac{1}{z_{12}\bar{z}_{34}}\big({\cal I}_{3}(z_{1},\bar{z}_{3})-{\cal I}_{3}(z_{2},\bar{z}_{3})-{\cal I}_{3}(z_{1},\bar{z}_{4})+{\cal I}_{3}(z_{2},\bar{z}_{4})\big)
\ea
\ee
and
\be
\ba
\label{eq:I1111z1312}
&{\cal I}_{1111}(z_{1},z_{3},\bar{z}_{1},\bar{z}_{2})
\\&
=\int d^{2}z\frac{1}{(z-z_{1})(\bar{z}-\bar{z}_{1})(z-z_{3})(\bar{z}-\bar{z}_{2})}
\\&
=\int d^{2}z\frac{1}{z_{13}\bar{z}_{12}}\big(\frac{1}{(z-z_{1})}-\frac{1}{(z-z_{3})}\big)\big(\frac{1}{(\bar{z}-\bar{z}_{1})}-\frac{1}{(\bar{z}-\bar{z}_{2})}\big)
\\&
=\int d^{2}z\frac{1}{z_{13}\bar{z}_{12}}\big(\frac{1}{(z-z_{1})(\bar{z}-\bar{z}_{1})}-\frac{1}{(z-z_{3})(\bar{z}-\bar{z}_{1})}-\frac{1}{(z-z_{1})(\bar{z}-\bar{z}_{2})}+\frac{1}{(z-z_{3})(\bar{z}-\bar{z}_{2})}\big)
\\&
=\frac{1}{z_{13}\bar{z}_{12}}\Big({\cal I}_{11}(z_{1},\bar{z}_{1})-{\cal I}_{3}(z_{3},\bar{z}_{1})-{\cal I}_{3}(z_{1},\bar{z}_{2})+{\cal I}_{3}(z_{3},\bar{z}_{2})\Big).
\ea
\ee
The other useful result is
\be
\ba
\label{eq:I11111}
&\quad {\cal I}_{11111}(z_{1},\bar{z}_{1},\bar{z}_{2},\bar{z}_{3},\bar{z}_{4})\\
&=\int d^{2}z\frac{1}{(z-z_{1})(\bar{z}-\bar{z}_{1})(\bar{z}-\bar{z}_{2})(\bar{z}-\bar{z}_{3})(\bar{z}-\bar{z}_{4})}\\
&=\frac{1}{\bar{z}_{12}\bar{z}_{34}}\int d^{2}z\frac{1}{(z-z_{1})}\Big(\frac{1}{(\bar{z}-\bar{z}_{1})}-\frac{1}{(\bar{z}-\bar{z}_{2})}\Big)\Big(\frac{1}{(\bar{z}-\bar{z}_{3})}-\frac{1}{(\bar{z}-\bar{z}_{4})}\Big)\\
&=\Big(\frac{1}{\bar{z}_{12}\bar{z}_{13}\bar{z}_{14}}{\cal I}_{11}(z_{1},\bar{z}_{1})-\frac{1}{\bar{z}_{12}\bar{z}_{23}\bar{z}_{24}}{\cal I}_{3}(z_{1},\bar{z}_{2})+\frac{1}{\bar{z}_{34}\bar{z}_{13}\bar{z}_{23}}{\cal I}_{3}(z_{1},\bar{z}_{3})-\frac{1}{\bar{z}_{34}\bar{z}_{14}\bar{z}_{24}}{\cal I}_{11}(z_{1},\bar{z}_{4})\Big).
\ea
\ee

By using $\frac{1}{(z-z_{a})(z-z_{b})}=\frac{1}{z_{ab}}\big(\frac{1}{(z-z_{a})}-\frac{1}{(z-z_{b})}\big)$ repeately, we find
\be
\ba
\label{eq:I11111111}
&{\cal I}_{11111111}(z_{1},z_{2},z_{3},z_{4},\bar{z}_{1},\bar{z}_{2},\bar{z}_{3},\bar{z}_{4})\\
=&\int d^{2}z\frac{1}{(z-z_{1})(z-z_{2})(z-z_{3})(z-z_{4})(\bar{z}-\bar{z}_{1})(\bar{z}-\bar{z}_{2})(\bar{z}-\bar{z}_{3})(\bar{z}-\bar{z}_{4})}\\
=&\frac{1}{z_{13}\bar{z}_{13}z_{24}\bar{z}_{24}}\Big({\cal I}_{1}(z_{1},z_{2})+{\cal I}_{1}(z_{2},z_{3})+{\cal I}_{1}(z_{1},z_{4})+{\cal I}_{1}(z_{3},z_{4})\\
&+{\cal I}_{1111}(z_{2},z_{3},\bar{z}_{1},\bar{z}_{4})+{\cal I}_{1111}(z_{1},z_{2},\bar{z}_{3},\bar{z}_{4})+{\cal I}_{1111}(z_{1},z_{4},\bar{z}_{2},\bar{z}_{3})+{\cal I}_{1111}(z_{3},z_{4},\bar{z}_{1},\bar{z}_{2})\\
&-{\cal I}_{1111}(z_{2},z_{1},\bar{z}_{2},\bar{z}_{3})-{\cal I}_{1111}(z_{1},z_{2},\bar{z}_{1},\bar{z}_{4})-{\cal I}_{1111}(z_{1},z_{4},\bar{z}_{1},\bar{z}_{2})-{\cal I}_{1111}(z_{4},z_{1},\bar{z}_{4},\bar{z}_{3})\\
&-{\cal I}_{1111}(z_{2},z_{3},\bar{z}_{2},\bar{z}_{1})-{\cal I}_{1111}(z_{3},z_{2},\bar{z}_{3},\bar{z}_{4})-{\cal I}_{1111}(z_{3},z_{4},\bar{z}_{3},\bar{z}_{2})-{\cal I}_{1111}(z_{3},z_{4},\bar{z}_{3},\bar{z}_{1})\Big).
\ea
\ee
Therefore, by using (\ref{eq:I1111z1234}) and (\ref{eq:I1111z1312}), we could express the complicated integral ${\cal I}_{11111111}$ in terms of  ${\cal I}_1$ and ${\cal I}_3$.

\section{Details of the integrals in R\'enyi entanglement entropy of excited state}
\label{app:ExEE}

In this appendix, let us show the details to evaluate the integral (\ref{eq:ExEE-next-order}). We show the following integral as an example\footnote{A similar integral can also be found in \cite{Chen:2018eqk}.}:
\be
\ba
&\int d^{2}z\frac{(z^{2}-L^{2})^{2}(\bar{z}^{2}-L^{2})^{2}}{4L^{6}|z|^{2}}\frac{1}{z^{2}}\frac{\bar{h}\bar{z}_{13}^{2}}{(\bar{z}-\bar{z}_{1})^{2}(\bar{z}-\bar{z}_{3})^{2}}\\
=&\frac{\bar{h}\bar{z}_{13}^{2}}{4L^{6}}\int_{0}^{\infty}d\rho\int_{0}^{2\pi}d\theta\rho\frac{(\rho^{2}e^{2i\theta}-L^{2})^{2}(\rho^{2}e^{-2i\theta}-L^{2})^{2}}{\rho^{4}e^{2i\theta}}\frac{1}{(\rho e^{-i\theta}-\bar{z}_{1})^{2}(\rho e^{-i\theta}-\bar{z}_{3})^{2}}.
\ea
\ee

Integrating on $\theta$, we find
\be
\ba
&\int d^{2}z\frac{(z^{2}-L^{2})^{2}(\bar{z}^{2}-L^{2})^{2}}{4L^{6}|z|^{2}}\frac{1}{z^{2}}\frac{\bar{h}\bar{z}_{13}^{2}}{(\bar{z}-\bar{z}_{1})^{2}(\bar{z}-\bar{z}_{3})^{2}}\\
=&\frac{\bar{h}\bar{z}_{13}^{2}}{4L^{6}}\int_{0}^{\infty}d\rho\frac{iL^{2}}{\bar{z}_{1}^{6}\rho^{5}}\Big\{2i\theta\rho^{4}[\bar{z}_{1}^{2}(L^{4}+\rho^{4})-L^{2}\rho^{4}]+(\bar{z}_{1}^{2}-L^{2})(\bar{z}_{1}^{4}-\rho^{4})(\bar{z}_{1}^{2}L^{2}-\rho^{4})\log(\bar{z}_{1}^{2}-e^{-2i\theta}\rho^{2})\Big\}\Big|_{\theta=0}^{2\pi},
\ea
\ee
where we used $e^{i0}=e^{i2\pi}$ and $z_3=-z_1$. Let us consider the progress where $\theta$ run from $0$ to $2\pi$. When $|\frac{\rho}{\bar{z}_{1}}|>1$, $e^{-2i\theta}\rho^{2}$ will go around $\bar{z}_1^2$ anti-clockwise twice, which means $\log(\bar{z}_{1}^{2}-e^{-2i\theta}\rho^{2})$ will contribute a factor $-4\pi i$. We thus find
\be
\ba
\label{eq:branch}
\log(\bar{z}_{1}^{2}-e^{-2i\theta}\rho^{2})&=\begin{cases}
0& |\frac{\rho^{2}}{\bar{z}_{1}^{2}}|<1\\
-4\pi i & |\frac{\rho^{2}}{\bar{z}_{1}^{2}}|>1
\end{cases}.
\ea
\ee
The integral thus becomes
\be
\ba
&\int d^{2}z\frac{(z^{2}-L^{2})^{2}(\bar{z}^{2}-L^{2})^{2}}{4L^{6}|z|^{2}}\frac{1}{z^{2}}\frac{\bar{h}\bar{z}_{13}^{2}}{(\bar{z}-\bar{z}_{1})^{2}(\bar{z}-\bar{z}_{3})^{2}}\\
=&-\frac{\bar{h}\pi\bar{z}_{13}^{2}}{L^{4}}\int_{0}^{\infty}d\rho\frac{1}{\bar{z}_{1}^{6}}[L^{4}\bar{z}_{1}^{2}\rho^{-1}+\rho^{3}\bar{z}_{1}^{2}-L^{2}\rho^{3}]
+\frac{\bar{h}\pi\bar{z}_{13}^{2}}{L^{6}}\int_{|z_{1}|}^{\infty}d\rho\frac{L^{2}(\bar{z}_{1}^{2}-L^{2})}{\bar{z}_{1}^{6}\rho^{5}}(\bar{z}_{1}^{4}-\rho^{4})(\bar{z}_{1}^{2}L^{2}-\rho^{4}).
\ea
\ee
Note that this integral divergent. We introduce the cutoff on $\rho$ as $(\frac{1}{\tilde{\Lambda}}, \Lambda)$.
\be
\ba
\xrightarrow{{\rm cutoff}}&-\frac{\bar{h}\pi\bar{z}_{13}^{2}}{L^{4}}\int_{\frac{1}{\tilde{\Lambda}}}^{\Lambda}d\rho\frac{1}{\bar{z}_{1}^{6}}[L^{4}\bar{z}_{1}^{2}\rho^{-1}+\rho^{3}\bar{z}_{1}^{2}-L^{2}\rho^{3}]
+\frac{\bar{h}\pi\bar{z}_{13}^{2}}{L^{6}}\int_{|z_{1}|}^{\Lambda}d\rho\frac{L^{2}(\bar{z}_{1}^{2}-L^{2})}{\bar{z}_{1}^{6}\rho^{5}}(\bar{z}_{1}^{4}-\rho^{4})(\bar{z}_{1}^{2}L^{2}-\rho^{4})\\
=&-\frac{4\pi\bar{h}\log(\tilde{\Lambda}|z_{1}|)}{\bar{z}_{1}^{2}}-\frac{4\pi\bar{h}\bar{z}_{1}^{2}\log(\Lambda/|z_{1}|)}{L^{4}}+\frac{\bar{h}\pi}{L^{4}\bar{z}_{1}^{4}}(\bar{z}_{1}^{2}-L^{2})(\bar{z}_{1}^{6}L^{2}|z_{1}|^{-4}-|z_{1}|^{4}).
\ea
\ee
The other integrals in (\ref{eq:ExEE-next-order}) can be evaluated in a similar way. We find the terms coupled with $\bar{T}$ becomes
 \be
 \ba
&\int d^{2}z\frac{|z^{2}-L^{2}|^{4}}{4L^{6}|z|^{2}}\frac{1}{z^{2}}\Big(\frac{\bar{h}\bar{z}_{13}^{2}}{(\bar{z}-\bar{z}_{1})^{2}(\bar{z}-\bar{z}_{3})^{2}}+\frac{\bar{h}\bar{z}_{24}^{2}}{(\bar{z}-\bar{z}_{2})^{2}(\bar{z}-\bar{z}_{4})^{2}}+\frac{\bar{z}_{23}\bar{z}_{14}}{\prod_{j=1}^{4}(\bar{z}-\bar{z}_{j})}\frac{\bar{\eta}\partial_{\bar{\eta}}G(\eta,\bar{\eta})}{G(\eta,\bar{\eta})}\Big)\\
=&-\frac{4\pi\bar{h}\log(\tilde{\Lambda}|z_{1}|)}{\bar{z}_{1}^{2}}-\frac{4\pi\bar{h}\bar{z}_{1}^{2}\log(\Lambda/|z_{1}|)}{L^{4}}+\frac{\bar{h}\pi}{L^{4}\bar{z}_{1}^{4}}(\bar{z}_{1}^{2}-L^{2})(\bar{z}_{1}^{6}L^{2}|z_{1}|^{-4}-|z_{1}|^{4})
\\&
-\frac{4\pi\bar{h}\log(\tilde{\Lambda}|z_{2}|)}{\bar{z}_{2}^{2}}-\frac{4\pi\bar{h}\bar{z}_{2}^{2}\log(\Lambda/|z_{2}|)}{L^{4}}+\frac{\bar{h}\pi}{L^{4}\bar{z}_{2}^{4}}(\bar{z}_{2}^{2}-L^{2})(\bar{z}_{2}^{6}L^{2}|z_{2}|^{-4}-|z_{2}|^{4})\\&+2\pi\frac{\bar{z}_{23}\bar{z}_{14}}{4L^{6}}\frac{\bar{\eta}\partial_{\bar{\eta}}G(\eta,\bar{\eta})}{G(\eta,\bar{\eta})}\Big\{\frac{4L^{4}\log(|z_{2}|/|z_{1}|)}{\bar{z}_{1}^{2}-\bar{z}_{2}^{2}}\\
&+\frac{2L^{6}\log(\tilde{\Lambda}|z_{1}|)}{\bar{z}_{1}^{2}\left(\bar{z}_{1}^{2}-\bar{z}_{2}^{2}\right)}-\frac{2L^{6}\log(\tilde{\Lambda|z_{2}|})}{\bar{z}_{2}^{2}\left(\bar{z}_{1}^{2}-\bar{z}_{2}^{2}\right)}-\frac{2\bar{z}_{1}^{2}L^{2}\log(\Lambda/|z_{1}|)}{\bar{z}_{1}^{2}-\bar{z}_{2}^{2}}+\frac{2\bar{z}_{2}^{2}L^{2}\log(\Lambda/|z_{2}|)}{\bar{z}_{1}^{2}-\bar{z}_{2}^{2}}\\
&
-\frac{(\bar{z}_{2}^{2}-L^{2})^{2}}{\bar{z}_{2}^{4}(\bar{z}_{1}^{2}-\bar{z}_{2}^{2})}(-\frac{1}{4}|\bar{z}_{2}|^{4}+\frac{L^{4}}{4}|\bar{z}_{2}|^{-4}\bar{z}_{2}^{4})+\frac{(\bar{z}_{1}^{2}-L^{2})^{2}}{\bar{z}_{1}^{4}(\bar{z}_{1}^{2}-\bar{z}_{2}^{2})}(-\frac{1}{4}|\bar{z}_{1}|^{4}+\frac{L^{4}}{4}|\bar{z}_{1}|^{-4}\bar{z}_{1}^{4})\Big\}.
 \ea
 \ee
The terms coupled with $T$ are found to be
 \be
 \ba
&\int d^{2}z\frac{(z^{2}-{L^{2}})^{2}(\bar{z}^{2}-{L^{2}})^{2}}{4L^{{6}}|z|^{2}}\frac{1}{\bar{z}^{2}}\Big(\frac{hz_{13}^{2}}{(z-z_{1})^{2}(z-z_{3})^{2}}+\frac{hz_{24}^{2}}{(z-z_{2})^{2}(z-z_{4})^{2}}+\frac{z_{23}z_{14}}{\prod_{j=1}^{4}(z-z_{j})}\frac{\eta\partial_{\eta}G(\eta,\bar{\eta})}{G(\eta,\bar{\eta})}\Big)\\
=&-\frac{4\pi hz_{1}^{2}\log(\Lambda/|z_{1}|)}{L^{4}}-\frac{4\pi h\log(|z_{1}|\tilde{\Lambda})}{z_{1}^{2}}+\frac{h\pi z_{1}^{2}}{L^{4}}\frac{1}{z_{1}^{6}}(L^{2}-z_{1}^{2})(-L^{2}z_{1}^{6}|z_{1}|^{-4}+|z_{1}|^{4})
\\&
-\frac{4\pi hz_{2}^{2}\log(\Lambda/|z_{2}|)}{L^{4}}-\frac{4\pi h\log(|z_{2}|\tilde{\Lambda})}{z_{2}^{2}}+\frac{h\pi}{L^{4}}\frac{1}{z_{2}^{4}}(L^{2}-z_{2}^{2})(-L^{2}z_{2}^{6}|z_{2}|^{-4}+|z_{2}|^{4})\\
&+\pi\frac{z_{23}z_{14}}{2L^{6}}\frac{\eta\partial_{\eta}G(\eta,\bar{\eta})}{G(\eta,\bar{\eta})}\Big\{\frac{4L^{4}\log(|z_{2}|/|z_{1}|)}{z_{1}^{2}-z_{2}^{2}}
\\&
+\frac{2L^{6}\log(\tilde{\Lambda}|z_{1}|)}{z_{1}^{2}\left(z_{1}^{2}-z_{2}^{2}\right)}-\frac{2L^{6}\log(\tilde{\Lambda}|z_{2}|)}{z_{2}^{2}\left(z_{1}^{2}-z_{2}^{2}\right)}-\frac{2L^{2}z_{1}^{2}\log(\Lambda/|z_{1}|)}{z_{1}^{2}-z_{2}^{2}}+\frac{2L^{2}z_{2}^{2}\log(\Lambda/|z_{2}|)}{z_{1}^{2}-z_{2}^{2}}\\
&+\frac{(L^{2}-z_{2}^{2})^{2}}{z_{2}^{4}(z_{1}^{2}-z_{2}^{2})}(\frac{1}{4}|z_{2}|^{4}-\frac{1}{4}L^{4}z_{2}^{4}|z_{2}|^{-4})-\frac{(L^{2}-z_{1}^{2})^{2}}{z_{1}^{4}(z_{1}^{2}-z_{2}^{2})}(\frac{1}{4}|z_{1}|^{4}-\frac{1}{4}L^{4}z_{1}^{4}|z_{1}|^{-4})\Big\}.
 \ea
 \ee
 Note that we have introduced a cutoff $\Lambda$ to regularize the divergence, which is different from the dimensional regularization.  Let us see the relationship with the dimensional regularization through an example
 \be
 {\cal I}_{3}(z_{1},\bar{z}_{2})=\int d\rho d\theta\frac{1}{(\rho e^{i\theta}-z_{1})(\rho e^{-i\theta}-\bar{z}_{2})}.
 \ee
By using (\ref{eq:branch}) and introducing the cutoff $\Lambda$, it is easy to evaluate ${\cal I}_{3}(z_1,\bar{z}_2)$ as
\be
{\cal I}_{3}(z_{1},\bar{z}_{2})=2\pi\log\Lambda-\pi\log|z_{12}|^{2}+{\cal O}(1/\Lambda^{2}).
\ee
Comparing with the result obtained by using dimensional regularization (\ref{eq:I3-regulated}), we find these two prescriptions of regularization are equivalent.

\section{Details of the integrals in OTOC}
\label{app:OTOC}

In this appendix, we show the details to evaluate the integral (\ref{eq:OTOC-Oc-G}). By using the Ward identity, the integral (\ref{eq:OTOC-Oc-G}) is simplified as
\be
\ba
&\lambda\frac{c}{24}(\frac{2\pi}{\beta})^{2}\Big\{
-\int d^{2}z_{a}|z_{a}|^{2}\frac{1}{\bar{z}_{a}^{2}}\frac{\langle T(z_{a})W(z_{1},\bar{z}_{1})W(z_{2},\bar{z}_{2})V(z_{3},\bar{z}_{3})V(z_{4},\bar{z}_{4})\rangle}{\langle W(z_{1},\bar{z}_{1})W(z_{2},\bar{z}_{2})V(z_{3},\bar{z}_{3})V(z_{4},\bar{z}_{4})\rangle}\\
&\qquad\qquad\quad -\lambda\int d^{2}z_{a}|z_{a}|^{2}\frac{1}{z_{a}^{2}}\frac{\langle\bar{T}(\bar{z}_{a})W(z_{1},\bar{z}_{1})W(z_{2},\bar{z}_{2})V(z_{3},\bar{z}_{3})V(z_{4},\bar{z}_{4})\rangle}{\langle W(z_{1},\bar{z}_{1})W(z_{2},\bar{z}_{2})V(z_{3},\bar{z}_{3})V(z_{4},\bar{z}_{4})\rangle}\\
&\qquad+ \int d^{2}z_{b}|z_{b}|^{2}\frac{1}{\bar{z}_{b}^{2}}\frac{\langle T(z_{b})W(z_{1},\bar{z}_{1})W(z_{2},\bar{z}_{2})\rangle}{\langle W(z_{1},\bar{z}_{1})W(z_{2},\bar{z}_{2})\rangle}+\int d^{2}z_{b}|z_{b}|^{2}\frac{1}{z_{b}^{2}}\frac{\langle\bar{T}(\bar{z}_{b})W(z_{1},\bar{z}_{1})W(z_{2},\bar{z}_{2})\rangle}{\langle W(z_{1},\bar{z}_{1})W(z_{2},\bar{z}_{2})\rangle}\\
&\qquad +\int d^{2}z_{c}|z_{c}|^{2}\frac{1}{\bar{z}_{c}^{2}}\frac{\langle\big(T(z_{c})\big)V(z_{3},\bar{z}_{3})V(z_{4},\bar{z}_{4})\rangle}{\langle V(z_{3},\bar{z}_{3})V(z_{4},\bar{z}_{4})\rangle}+\int d^{2}z_{c}|z_{c}|^{2}\frac{1}{z_{c}^{2}}\frac{\langle\bar{T}(\bar{z}_{c})V(z_{3},\bar{z}_{3})V(z_{4},\bar{z}_{4})\rangle}{\langle V(z_{3},\bar{z}_{3})V(z_{4},\bar{z}_{4})\rangle}\Big\}\\
=&-\int d^{2}z_{a}|z_{a}|^{2}\frac{1}{\bar{z}_{a}^{2}}\Big(\frac{z_{14}z_{23}}{\prod_{i=1}^{4}(z-z_{i})}\frac{\eta\partial_{\eta}G(\eta,\bar{\eta})}{G(\eta,\bar{\eta})}\Big)-\int d^{2}z_{a}|z_{a}|^{2}\frac{1}{z_{a}^{2}}\Big(\frac{\bar{z}_{14}\bar{z}_{23}}{\prod_{i=1}^{4}(\bar{z}-\bar{z}_{i})}\frac{\bar{\eta}\partial_{\bar{\eta}}G(\eta,\bar{\eta})}{G(\eta,\bar{\eta})}\Big).
\ea
\ee

Let us evaluate this integral step by step. We first consider
\be
\ba
&\int d^{2}z_{a}|z_{a}|^{2}\frac{1}{\bar{z}_{a}^{2}}\frac{z_{14}z_{23}}{\prod_{i=1}^{4}(z-z_{i})}\frac{\eta\partial_{\eta}G(\eta,\bar{\eta})}{G(\eta,\bar{\eta})}\\
=&\frac{\eta\partial_{\eta}G(\eta,\bar{\eta})}{G(\eta,\bar{\eta})}\int_{0}^{\infty}d\rho\int_{0}^{2\pi}d\theta\rho^{3}\frac{1}{\rho^{2}e^{-2i\theta}}\frac{z_{14}z_{23}}{\prod_{i=1}^{4}(\rho e^{i\theta}-z_{i})}\\
=&-\frac{\eta\partial_{\eta}G(\eta,\bar{\eta})}{G(\eta,\bar{\eta})}iz_{14}z_{23}\int_{0}^{\infty}d\rho\frac{\Big(\frac{z_{1}\log(z_{1}-e^{i\theta}\rho)}{z_{12}z_{13}z_{14}}+\frac{z_{3}\log(-z_{3}+e^{i\theta}\rho)}{z_{13}z_{23}z_{34}}-\frac{z_{4}\log(-z_{4}+e^{i\theta}\rho)}{z_{14}z_{24}z_{34}}-\frac{z_{2}\log(-z_{2}+e^{i\theta}\rho)}{z_{12}z_{23}z_{24}}\Big)\Big|_{\theta=0}^{2\pi}}{\rho}\\
\xrightarrow{\rm cut-off}&2\pi\frac{\eta\partial_{\eta}G(\eta,\bar{\eta})}{G(\eta,\bar{\eta})}z_{14}z_{23}\Big(\frac{z_{1}}{z_{12}z_{13}z_{14}}\log\frac{1}{|z_{1}|}+\frac{z_{3}}{z_{13}z_{23}z_{34}}\log\frac{1}{|z_{3}|}-\frac{z_{4}}{z_{14}z_{24}z_{34}}\log\frac{1}{|z_{4}|}-\frac{z_{2}}{z_{12}z_{23}z_{24}}\log\frac{1}{|z_{2}|}\Big).
\ea
\ee
The other integral is evaluated as
\be
\ba
&\int d^{2}z_{a}|z_{a}|^{2}\frac{1}{z_{a}^{2}}\frac{\bar{z}_{14}\bar{z}_{23}}{\prod_{i=1}^{4}(\bar{z}-\bar{z}_{i})}\frac{\bar{\eta}\partial_{\bar{\eta}}G(\eta,\bar{\eta})}{G(\eta,\bar{\eta})}\\
=&-\frac{\bar{\eta}\partial_{\bar{\eta}}G(\eta,\bar{\eta})}{G(\eta,\bar{\eta})}\bar{z}_{14}\bar{z}_{23}2\pi\\
&\Big\{\frac{\bar{z}_{1}}{\bar{z}_{12}\bar{z}_{13}\bar{z}_{14}}\log(|z_{1}|)-\frac{\bar{z}_{2}}{\bar{z}_{12}\bar{z}_{23}\bar{z}_{24}}\log(|z_{2}|)+\frac{\bar{z}_{3}}{\bar{z}_{13}\bar{z}_{23}\bar{z}_{34}}\log(|z_{3}|)-\frac{\bar{z}_{4}}{\bar{z}_{14}\bar{z}_{24}\bar{z}_{34}}\log(|z_{4}|)\Big\}.
\ea
\ee

In summary, (\ref{eq:OTOC-next-order}) is evaluated as
\be
\ba
&-\int d^{2}z_{a}|z_{a}|^{2}\frac{1}{\bar{z}_{a}^{2}}\Big(\frac{z_{14}z_{23}}{\prod_{i=1}^{4}(z-z_{i})}\frac{\eta\partial_{\eta}G(\eta,\eta)}{G(\eta,\eta)}\Big)-\int d^{2}z_{a}|z_{a}|^{2}\frac{1}{z_{a}^{2}}\Big(\frac{\bar{z}_{14}\bar{z}_{23}}{\prod_{i=1}^{4}(\bar{z}-\bar{z}_{i})}\frac{\bar{\eta}\partial_{\bar{\eta}}G(\eta,\eta)}{G(\eta,\eta)}\Big)\\
=&-2\pi\frac{\eta\partial_{\eta}G(\eta,\eta)}{G(\eta,\eta)}z_{14}z_{23}\Big(\frac{z_{1}}{z_{12}z_{13}z_{14}}\log\frac{1}{|z_{1}|}+\frac{z_{3}}{z_{13}z_{23}z_{34}}\log\frac{1}{|z_{3}|}-\frac{z_{4}}{z_{14}z_{24}z_{34}}\log\frac{1}{|z_{4}|}-\frac{z_{2}}{z_{12}z_{23}z_{24}}\log\frac{1}{|z_{2}|}\Big)\\
&+2\pi\frac{\bar{\eta}\partial_{\bar{\eta}}G(\eta,\eta)}{G(\eta,\eta)}\bar{z}_{14}\bar{z}_{23}\Big(\frac{\bar{z}_{1}}{\bar{z}_{12}\bar{z}_{13}\bar{z}_{14}}\log|z_{1}|-\frac{\bar{z}_{2}}{\bar{z}_{12}\bar{z}_{23}\bar{z}_{24}}\log|z_{2}|+\frac{\bar{z}_{3}}{\bar{z}_{13}\bar{z}_{23}\bar{z}_{34}}\log|z_{3}|-\frac{\bar{z}_{4}}{\bar{z}_{14}\bar{z}_{24}\bar{z}_{34}}\log|z_{4}|\Big).
\ea
\ee


\begin{thebibliography}{99}



\bibitem{Smirnov:2016lqw}
  F.~A.~Smirnov and A.~B.~Zamolodchikov,
 ``On space of integrable quantum field theories,''
  Nucl.\ Phys.\ B {\bf 915}, 363 (2017)
  [arXiv:1608.05499 [hep-th]].





\bibitem{Zamolodchikov:2004ce}
  A.~B.~Zamolodchikov,
 ``Expectation value of composite field T anti-T in two-dimensional quantum field theory,''
  hep-th/0401146.

\bibitem{Cavaglia:2016oda}
  A.~Cavaglia, S.~Negro, I.~M.~Szecsenyi and R.~Tateo,
 ``$T \bar{T}$-deformed 2D Quantum Field Theories,''
  JHEP {\bf 1610}, 112 (2016)
  [arXiv:1608.05534 [hep-th]].


\bibitem{Dubovsky:2012wk}
  S.~Dubovsky, R.~Flauger and V.~Gorbenko,
 ``Solving the Simplest Theory of Quantum Gravity,''
  JHEP {\bf 1209}, 133 (2012)
  [arXiv:1205.6805 [hep-th]].


\bibitem{Caselle:2013dra}
  M.~Caselle, D.~Fioravanti, F.~Gliozzi and R.~Tateo,
 ``Quantisation of the effective string with TBA,''
  JHEP {\bf 1307}, 071 (2013)
  [arXiv:1305.1278 [hep-th]].


\bibitem{Dubovsky:2017cnj}
  S.~Dubovsky, V.~Gorbenko and M.~Mirbabayi,
 ``Asymptotic fragility, near AdS$_{2}$ holography and $ T\overline{T} $,''
  JHEP {\bf 1709}, 136 (2017)
  [arXiv:1706.06604 [hep-th]].


\bibitem{Cardy:2018sdv}
  J.~Cardy,
 ``The $ T\overline{T} $ deformation of quantum field theory as random geometry,''
  JHEP {\bf 1810}, 186 (2018)
  [arXiv:1801.06895 [hep-th]].



\bibitem{Aharony:2018vux}
  O.~Aharony and T.~Vaknin,
 ``The TT* deformation at large central charge,''
  JHEP {\bf 1805}, 166 (2018)
  [arXiv:1803.00100 [hep-th]].


\bibitem{Bonelli:2018kik}
  G.~Bonelli, N.~Doroud and M.~Zhu,
 ``$T \bar{T}$-deformations in closed form,''
  JHEP {\bf 1806}, 149 (2018)
  [arXiv:1804.10967 [hep-th]].


\bibitem{Chakraborty:2018kpr}
  S.~Chakraborty, A.~Giveon, N.~Itzhaki and D.~Kutasov,
  ``Entanglement beyond AdS,''
  Nucl.\ Phys.\ B {\bf 935}, 290 (2018)
  [arXiv:1805.06286 [hep-th]].



\bibitem{Datta:2018thy}
  S.~Datta and Y.~Jiang,
 ``$T\bar{T}$-deformed partition functions,''
  JHEP {\bf 1808}, 106 (2018)
  [arXiv:1806.07426 [hep-th]].


\bibitem{Aharony:2018bad}
  O.~Aharony, S.~Datta, A.~Giveon, Y.~Jiang and D.~Kutasov,
 ``Modular invariance and uniqueness of $T\bar{T}$-deformed CFT,''
  JHEP {\bf 1901}, 086 (2019)
  [arXiv:1808.02492 [hep-th]].


\bibitem{Chen:2018eqk}
  B.~Chen, L.~Chen and P.~X.~Hao,
 ``Entanglement entropy in $T\overline{T}$-deformed CFT,''
  Phys.\ Rev.\ D {\bf 98}, no. 8, 086025 (2018)
  [arXiv:1807.08293 [hep-th]].


\bibitem{Conti:2018tca}
  R.~Conti, S.~Negro and R.~Tateo,
 ``The $ \mathrm{T}\overline{\mathrm{T}} $ perturbation and its geometric interpretation,''
  JHEP {\bf 1902}, 085 (2019)
  [arXiv:1809.09593 [hep-th]].



\bibitem{Lashkevich:2018jmo}
  M.~Lashkevich and Y.~Pugai,
 ``The complex sinh-Gordon model: form factors of descendant operators and current-current perturbations,''
  JHEP {\bf 1901}, 071 (2019)
  [arXiv:1811.02631 [hep-th]].


\bibitem{Chang:2018dge}
  C.~K.~Chang, C.~Ferko and S.~Sethi,
 ``Supersymmetry and $ T\overline{T} $ deformations,''
  JHEP {\bf 1904}, 131 (2019)
  [arXiv:1811.01895 [hep-th]].


\bibitem{Baggio:2018rpv}
  M.~Baggio, A.~Sfondrini, G.~Tartaglino-Mazzucchelli and H.~Walsh,
 ``On $ T\overline{T} $ deformations and supersymmetry,''
  JHEP {\bf 1906}, 063 (2019)
  [arXiv:1811.00533 [hep-th]].



\bibitem{Dubovsky:2018bmo}
  S.~Dubovsky, V.~Gorbenko and G.~Hernandez-Chifflet,
 ``$ T\overline{T} $ partition function from topological gravity,''
  JHEP {\bf 1809}, 158 (2018)
  [arXiv:1805.07386 [hep-th]].


\bibitem{Conti:2018jho}
  R.~Conti, L.~Iannella, S.~Negro and R.~Tateo,
 ``Generalised Born-Infeld models, Lax operators and the $ \mathrm{T}\overline{\mathrm{T}} $ perturbation,''
  JHEP {\bf 1811}, 007 (2018)
  [arXiv:1806.11515 [hep-th]].


\bibitem{Chen:2018keo}
  C.~Chen, P.~Conkey, S.~Dubovsky and G.~Hernandez-Chifflet,
 ``Undressing Confining Flux Tubes with $T\bar T$,''
  Phys.\ Rev.\ D {\bf 98}, no. 11, 114024 (2018)
  [arXiv:1808.01339 [hep-th]].


\bibitem{Santilli:2018xux}
  L.~Santilli and M.~Tierz,
 ``Large N phase transition in $ T\overline{T} $ -deformed 2d Yang-Mills theory on the sphere,''
  JHEP {\bf 1901}, 054 (2019)
  [arXiv:1810.05404 [hep-th]].


\bibitem{Jiang:2019tcq}
  Y.~Jiang,
 ``Expectation value of $\mathrm{T}\overline{\mathrm{T}}$ operator in curved spacetimes,''
  arXiv:1903.07561 [hep-th].


\bibitem{LeFloch:2019rut}
  B.~Le Floch and M.~Mezei,
 ``Solving a family of $T\bar{T}$-like theories,''
  arXiv:1903.07606 [hep-th].


\bibitem{Jiang:2019hux}
  H.~Jiang, A.~Sfondrini and G.~Tartaglino-Mazzucchelli,
 ``$T\bar{T}$-deformations with $\mathcal{N}=(0,2)$ supersymmetry,''
  arXiv:1904.04760 [hep-th].


\bibitem{Conti:2019dxg}
  R.~Conti, S.~Negro and R.~Tateo,
 ``Conserved currents and $\text{T}\bar{\text{T}}_s$ irrelevant deformations of 2D integrable field theories,''
  arXiv:1904.09141 [hep-th].


\bibitem{Chang:2019kiu}
  C.~K.~Chang, C.~Ferko, S.~Sethi, A.~Sfondrini and G.~Tartaglino-Mazzucchelli,
 ``$T\bar{T}$ Flows and (2,2) Supersymmetry,''
  arXiv:1906.00467 [hep-th].



\bibitem{Cardy:2018jho}
  J.~Cardy,
 ``$T\overline T$ deformations of non-Lorentz invariant field theories,''
  arXiv:1809.07849 [hep-th].



\bibitem{Guica:2017lia}
  M.~Guica,
 ``An integrable Lorentz-breaking deformation of two-dimensional CFTs,''
  SciPost Phys.\  {\bf 5}, no. 5, 048 (2018)
  [arXiv:1710.08415 [hep-th]].


\bibitem{Chakraborty:2018vja}
  S.~Chakraborty, A.~Giveon and D.~Kutasov,
 ``$ J\overline{T} $ deformed CFT$_{2}$ and string theory,''
  JHEP {\bf 1810}, 057 (2018)
  [arXiv:1806.09667 [hep-th]].


\bibitem{Aharony:2018ics}
  O.~Aharony, S.~Datta, A.~Giveon, Y.~Jiang and D.~Kutasov,
 ``Modular covariance and uniqueness of $J\bar{T}$-deformed CFTs,''
  JHEP {\bf 1901}, 085 (2019)
  [arXiv:1808.08978 [hep-th]].


\bibitem{Nakayama:2018ujt}
  Y.~Nakayama,
 ``Very Special $T\bar{J}$ deformed CFT,''
  Phys.\ Rev.\ D {\bf 99}, no. 8, 085008 (2019)
  [arXiv:1811.02173 [hep-th]].


\bibitem{Guica:2019vnb}
  M.~Guica,
 ``On correlation functions in $J\bar T$-deformed CFTs,''
  J.\ Phys.\ A {\bf 52}, no. 18, 184003 (2019)
  [arXiv:1902.01434 [hep-th]].


\bibitem{Giveon:2017nie}
  A.~Giveon, N.~Itzhaki and D.~Kutasov,
 ``$ \mathrm{T}\overline{\mathrm{T}} $ and LST,''
  JHEP {\bf 1707}, 122 (2017)
  [arXiv:1701.05576 [hep-th]].


\bibitem{Giveon:2017myj}
  A.~Giveon, N.~Itzhaki and D.~Kutasov,
 ``A solvable irrelevant deformation of AdS$_{3}$/CFT$_{2}$,''
  JHEP {\bf 1712}, 155 (2017)
  [arXiv:1707.05800 [hep-th]].


\bibitem{Asrat:2017tzd}
  M.~Asrat, A.~Giveon, N.~Itzhaki and D.~Kutasov,
 ``Holography Beyond AdS,''
  Nucl.\ Phys.\ B {\bf 932}, 241 (2018)
  [arXiv:1711.02690 [hep-th]].

\bibitem{Giribet:2017imm}
  G.~Giribet,
 ``$T\bar{T}$-deformations, AdS/CFT and correlation functions,''
  JHEP {\bf 1802}, 114 (2018)
  [arXiv:1711.02716 [hep-th]].


\bibitem{Baggio:2018gct}
  M.~Baggio and A.~Sfondrini,
 ``Strings on NS-NS Backgrounds as Integrable Deformations,''
  Phys.\ Rev.\ D {\bf 98}, no. 2, 021902 (2018)
  [arXiv:1804.01998 [hep-th]].





\bibitem{Apolo:2018qpq}
  L.~Apolo and W.~Song,
 ``Strings on warped AdS$_{3}$ via $ \mathrm{T}\bar{\mathrm{J}} $ deformations,''
  JHEP {\bf 1810}, 165 (2018)
  [arXiv:1806.10127 [hep-th]].


\bibitem{Babaro:2018cmq}
  J.~P.~Babaro, V.~F.~Foit, G.~Giribet and M.~Leoni,
 ``$ T\overline{T} $ type deformation in the presence of a boundary,''
  JHEP {\bf 1808}, 096 (2018)
  [arXiv:1806.10713 [hep-th]].


\bibitem{Chakraborty:2018aji}
  S.~Chakraborty,
 ``Wilson loop in a $T\bar{T}$ like deformed $\rm{CFT}_2$,''
  Nucl.\ Phys.\ B {\bf 938}, 605 (2019)
  [arXiv:1809.01915 [hep-th]].


\bibitem{Araujo:2018rho}
  T.~Araujo, E.~O.~Colgain, Y.~Sakatani, M.~M.~Sheikh-Jabbari and H.~Yavartanoo,
 ``Holographic integration of $T \bar{T}$ \& $J \bar{T}$ via $O(d,d)$,''
  JHEP {\bf 1903}, 168 (2019)
  [arXiv:1811.03050 [hep-th]].


\bibitem{Giveon:2019fgr}
  A.~Giveon,
 ``Comments on $T\bar T$, $J\bar{T}$ and String Theory,''
  arXiv:1903.06883 [hep-th].


\bibitem{Chakraborty:2019mdf}
  S.~Chakraborty, A.~Giveon and D.~Kutasov,
 ``$T\bar{T}$, $J\bar{T}$, $T\bar{J}$ and String Theory,''
  arXiv:1905.00051 [hep-th].


\bibitem{Nakayama:2019mvq}
  Y.~Nakayama,
 ``Holographic Dual of Conformal Field Theories with Very Special $T\bar{J}$ Deformations,''
  arXiv:1905.05353 [hep-th].


\bibitem{McGough:2016lol}
  L.~McGough, M.~Mezei and H.~Verlinde,
 ``Moving the CFT into the bulk with $ T\overline{T} $,''
  JHEP {\bf 1804}, 010 (2018)
  [arXiv:1611.03470 [hep-th]].


\bibitem{Shyam:2017znq}
  V.~Shyam,
 ``Background independent holographic dual to $T\bar{T}$-deformed CFT with large central charge in 2 dimensions,''
  JHEP {\bf 1710}, 108 (2017)
  [arXiv:1707.08118 [hep-th]].


\bibitem{Kraus:2018xrn}
  P.~Kraus, J.~Liu and D.~Marolf,
 ``Cutoff AdS$_{3}$ versus the $ T\overline{T} $ deformation,''
  JHEP {\bf 1807}, 027 (2018)
  [arXiv:1801.02714 [hep-th]].


\bibitem{Cottrell:2018skz}
  W.~Cottrell and A.~Hashimoto,
 ``Comments on $T \bar T$ double trace deformations and boundary conditions,''
  Phys.\ Lett.\ B {\bf 789}, 251 (2019)
  [arXiv:1801.09708 [hep-th]].


\bibitem{Bzowski:2018pcy}
  A.~Bzowski and M.~Guica,
 ``The holographic interpretation of $J \bar T$-deformed CFTs,''
  JHEP {\bf 1901}, 198 (2019)
  [arXiv:1803.09753 [hep-th]].


\bibitem{Taylor:2018xcy}
  M.~Taylor,
 ``TT deformations in general dimensions,''
  arXiv:1805.10287 [hep-th].


\bibitem{Hartman:2018tkw}
  T.~Hartman, J.~Kruthoff, E.~Shaghoulian and A.~Tajdini,
 ``Holography at finite cutoff with a $T^2$ deformation,''
  JHEP {\bf 1903}, 004 (2019)
  [arXiv:1807.11401 [hep-th]].


\bibitem{Shyam:2018sro}
  V.~Shyam,
 ``Finite Cutoff AdS$_{5}$ Holography and the Generalized Gradient Flow,''
  JHEP {\bf 1812}, 086 (2018)
  [arXiv:1808.07760 [hep-th]].


\bibitem{Caputa:2019pam}
  P.~Caputa, S.~Datta and V.~Shyam,
 ``Sphere partition functions \& cut-off AdS,''
  JHEP {\bf 1905}, 112 (2019)
  [arXiv:1902.10893 [hep-th]].


\bibitem{Gorbenko:2018oov}
  V.~Gorbenko, E.~Silverstein and G.~Torroba,
 ``dS/dS and $ T\overline{T} $,''
  JHEP {\bf 1903}, 085 (2019)
  [arXiv:1811.07965 [hep-th]].


\bibitem{Apolo:2019yfj}
  L.~Apolo and W.~Song,
 ``Heating up holography for single-trace $J\bar{T}$-deformations,''
  arXiv:1907.03745 [hep-th].


\bibitem{Calabrese:2004eu}
  P.~Calabrese and J.~L.~Cardy,
 ``Entanglement entropy and quantum field theory,''
  J.\ Stat.\ Mech.\  {\bf 0406}, P06002 (2004)
  [hep-th/0405152].


\bibitem{Kitaev:2005dm}
  A.~Kitaev and J.~Preskill,
 ``Topological entanglement entropy,''
  Phys.\ Rev.\ Lett.\  {\bf 96}, 110404 (2006)
  [hep-th/0510092].


\bibitem{Levin:2006zz}
  M.~Levin and X.~G.~Wen,
 ``Detecting Topological Order in a Ground State Wave Function,''
  Phys.\ Rev.\ Lett.\  {\bf 96}, 110405 (2006)
  [cond-mat/0510613 [cond-mat.str-el]].


\bibitem{Nozaki:2014hna}
  M.~Nozaki, T.~Numasawa and T.~Takayanagi,
 ``Quantum Entanglement of Local Operators in Conformal Field Theories,''
  Phys.\ Rev.\ Lett.\  {\bf 112}, 111602 (2014)
  [arXiv:1401.0539 [hep-th]].


\bibitem{He:2014mwa}
  S.~He, T.~Numasawa, T.~Takayanagi and K.~Watanabe,
 ``Quantum dimension as entanglement entropy in two dimensional conformal field theories,''
  Phys.\ Rev.\ D {\bf 90}, no. 4, 041701 (2014)
  [arXiv:1403.0702 [hep-th]].


\bibitem{Guo:2015uwa}
  W.~Z.~Guo and S.~He,
 ``R\'enyi entropy of locally excited states with thermal and boundary effect in 2D CFTs,''
  JHEP {\bf 1504}, 099 (2015)
  [arXiv:1501.00757 [hep-th]].


\bibitem{Chen:2015usa}
  B.~Chen, W.~Z.~Guo, S.~He and J.~q.~Wu,
 ``Entanglement Entropy for Descendent Local Operators in 2D CFTs,''
  JHEP {\bf 1510}, 173 (2015)
  [arXiv:1507.01157 [hep-th]].

\bibitem{He:2017vyf}
  S.~He, F.~L.~Lin and J.~j.~Zhang,
 ``Subsystem eigenstate thermalization hypothesis for entanglement entropy in CFT,''
  JHEP {\bf 1708}, 126 (2017)
  [arXiv:1703.08724 [hep-th]].

\bibitem{He:2017txy}
  S.~He, F.~L.~Lin and J.~j.~Zhang,
``Dissimilarities of reduced density matrices and eigenstate thermalization hypothesis,''
  JHEP {\bf 1712}, 073 (2017)
  [arXiv:1708.05090 [hep-th]].

\bibitem{He:2017lrg}
  S.~He,
 ``Conformal bootstrap to R\'enyi entropy in 2D Liouville and super-Liouville CFTs,''
  Phys.\ Rev.\ D {\bf 99}, no. 2, 026005 (2019)
  [arXiv:1711.00624 [hep-th]].


\bibitem{Guo:2018lqq}
  W.~Z.~Guo, S.~He and Z.~X.~Luo,
 ``Entanglement entropy in (1+1)D CFTs with multiple local excitations,''
  JHEP {\bf 1805}, 154 (2018)
  [arXiv:1802.08815 [hep-th]].


\bibitem{Apolo:2018oqv}
  L.~Apolo, S.~He, W.~Song, J.~Xu and J.~Zheng,
 ``Entanglement and chaos in warped conformal field theories,''
  JHEP {\bf 1904}, 009 (2019)
  [arXiv:1812.10456 [hep-th]].


\bibitem{Caputa:2014vaa}
  P.~Caputa, M.~Nozaki and T.~Takayanagi,
 ``Entanglement of local operators in large-N conformal field theories,''
  PTEP {\bf 2014}, 093B06 (2014)
  [arXiv:1405.5946 [hep-th]].


\bibitem{Caputa:2014eta}
  P.~Caputa, J.~Simon, A.~Stikonas and T.~Takayanagi,
 ``Quantum Entanglement of Localized Excited States at Finite Temperature,''
  JHEP {\bf 1501}, 102 (2015)
  [arXiv:1410.2287 [hep-th]].


\bibitem{Jahn:2017xsg}
  A.~Jahn and T.~Takayanagi,
 ``Holographic entanglement entropy of local quenches in AdS4/CFT3: a finite-element approach,''
  J.\ Phys.\ A {\bf 51}, no. 1, 015401 (2018)
  [arXiv:1705.04705 [hep-th]].


\bibitem{Miyaji:2018atq}
  M.~Miyaji,
 ``Time Evolution after Double Trace Deformation,''
  JHEP {\bf 1810}, 074 (2018)
  [arXiv:1806.10807 [hep-th]].


\bibitem{Kusuki:2018wpa}
  Y.~Kusuki,
 ``Light Cone Bootstrap in General 2D CFTs and Entanglement from Light Cone Singularity,''
  JHEP {\bf 1901}, 025 (2019)
  [arXiv:1810.01335 [hep-th]].


\bibitem{Shimaji:2018czt}
  T.~Shimaji, T.~Takayanagi and Z.~Wei,
 ``Holographic Quantum Circuits from Splitting/Joining Local Quenches,''
  JHEP {\bf 1903}, 165 (2019)
  [arXiv:1812.01176 [hep-th]].


\bibitem{Kusuki:2019gjs}
  Y.~Kusuki and M.~Miyaji,
 ``Entanglement Entropy, OTOC and Bootstrap in 2D CFTs from Regge and Light Cone Limits of Multi-point Conformal Block,''
  arXiv:1905.02191 [hep-th].


\bibitem{Caputa:2019avh}
  P.~Caputa, T.~Numasawa, T.~Shimaji, T.~Takayanagi and Z.~Wei,
 ``Double Local Quenches in 2D CFTs and Gravitational Force,''
  arXiv:1905.08265 [hep-th].



\bibitem{Donnelly:2018bef}
  W.~Donnelly and V.~Shyam,
 ``Entanglement entropy and $T \overline{T}$ deformation,''
  Phys.\ Rev.\ Lett.\  {\bf 121}, no. 13, 131602 (2018)
  [arXiv:1806.07444 [hep-th]].


\bibitem{Park:2018snf}
  C.~Park,
 ``Holographic Entanglement Entropy in Cutoff AdS,''
  Int.\ J.\ Mod.\ Phys.\ A {\bf 33}, no. 36, 1850226 (2019)
  [arXiv:1812.00545 [hep-th]].


\bibitem{Banerjee:2019ewu}
  A.~Banerjee, A.~Bhattacharyya and S.~Chakraborty,
 ``Entanglement Entropy for $TT$ deformed CFT in general dimensions,''
  arXiv:1904.00716 [hep-th].


\bibitem{Murdia:2019fax}
  C.~Murdia, Y.~Nomura, P.~Rath and N.~Salzetta,
 ``Comments on Holographic Entanglement Entropy in $TT$ Deformed CFTs,''
  Phys.\ Rev.\ D {\bf 100}, 026011 (2019)
  [arXiv:1904.04408 [hep-th]].


\bibitem{Ota:2019yfe}
  T.~Ota,
 ``Comments on holographic entanglements in cutoff AdS,''
  arXiv:1904.06930 [hep-th].


\bibitem{Sun:2019ijq}
  Y.~Sun and J.~R.~Sun,
 ``Note on the R\'enyi entropy of 2D perturbed fermions,''
  Phys.\ Rev.\ D {\bf 99}, no. 10, 106008 (2019)
  [arXiv:1901.08796 [hep-th]].


\bibitem{Jeong:2019ylz}
  H.~S.~Jeong, K.~Y.~Kim and M.~Nishida,
 ``Entanglement and R\'enyi Entropy of Multiple Intervals in $T\overline{T}$-Deformed CFT and Holography,''
  arXiv:1906.03894 [hep-th].




\bibitem{Cardy:2019qao}
  J.~Cardy,
 ``$T\overline T$ deformation of correlation functions,''
  arXiv:1907.03394 [hep-th].


\bibitem{Shenker:2014cwa}
  S.~H.~Shenker and D.~Stanford,
 ``Stringy effects in scrambling,''
  JHEP {\bf 1505}, 132 (2015)
  [arXiv:1412.6087 [hep-th]].


\bibitem{Maldacena:2015waa}
  J.~Maldacena, S.~H.~Shenker and D.~Stanford,
 ``A bound on chaos,''
  JHEP {\bf 1608}, 106 (2016)
  [arXiv:1503.01409 [hep-th]].


\bibitem{Roberts:2014ifa}
  D.~A.~Roberts and D.~Stanford,
 ``Two-dimensional conformal field theory and the butterfly effect,''
  Phys.\ Rev.\ Lett.\  {\bf 115}, no. 13, 131603 (2015)
  [arXiv:1412.5123 [hep-th]].


\bibitem{Belavin:1984vu}
  A.~A.~Belavin, A.~M.~Polyakov and A.~B.~Zamolodchikov,
 ``Infinite Conformal Symmetry in Two-Dimensional Quantum Field Theory,''
  Nucl.\ Phys.\ B {\bf 241}, 333 (1984).


\bibitem{Moore:1988uz}
  G.~W.~Moore and N.~Seiberg,
 ``Polynomial Equations for Rational Conformal Field Theories,''
  Phys.\ Lett.\ B {\bf 212}, 451 (1988).


\bibitem{Moore:1988ss}
  G.~W.~Moore and N.~Seiberg,
 ``Naturality in Conformal Field Theory,''
  Nucl.\ Phys.\ B {\bf 313}, 16 (1989).


\bibitem{Dolan:2000ut}
  F.~A.~Dolan and H.~Osborn,
 ``Conformal four-point functions and the operator product expansion,''
  Nucl.\ Phys.\ B {\bf 599}, 459 (2001)
  [hep-th/0011040].


\bibitem{Perlmutter:2016pkf}
  E.~Perlmutter,
 ``Bounding the Space of Holographic CFTs with Chaos,''
  JHEP {\bf 1610}, 069 (2016)
  [arXiv:1602.08272 [hep-th]].


\bibitem{Fitzpatrick:2014vua}
  A.~L.~Fitzpatrick, J.~Kaplan and M.~T.~Walters,
  ``Universality of Long-Distance AdS Physics from the CFT Bootstrap,''
  JHEP {\bf 1408}, 145 (2014)
  [arXiv:1403.6829 [hep-th]].



\bibitem{Gross:2019ach}
  D.~J.~Gross, J.~Kruthoff, A.~Rolph and E.~Shaghoulian,
 ``$T\overline{T}$ in AdS$_2$ and Quantum Mechanics,''
  arXiv:1907.04873 [hep-th].



\end{thebibliography}
\end{document}